\documentclass[preprints,article,accept,moreauthors,pdftex]{Definitions/mdpi}

\setitemize{parsep=6pt,itemsep=0pt,leftmargin=*,labelsep=5.5mm}
\setenumerate{parsep=6pt,itemsep=0pt,leftmargin=*,labelsep=5.5mm}
\setlist[description]{itemsep=0mm}

\usepackage[labelformat=simple]{subfig}

\firstpage{1} 
\pubvolume{xx}
\issuenum{1}
\articlenumber{5}
\pubyear{2020}
\copyrightyear{2020}
\history{Received: 17 March 2020; Accepted: 21 April 2020; Published: date}

\usepackage{svg}
\usepackage[english]{babel}
\usepackage{mathtools, cuted}
\usepackage{oubraces}
\usepackage{pdflscape}
\usepackage{colortbl}
\usepackage{tabularx}
\usepackage{array}
\usepackage{booktabs}
\usepackage{multirow}
\usepackage{graphicx}
\usepackage{ragged2e}
\usepackage{rotating}
\usepackage{xcolor}
\usepackage{natbib}
\usepackage{interval}
\usepackage[font=scriptsize,labelfont=bf]{caption}
\newlist{mylist}{enumerate*}{1}
\setlist[mylist]{label=(\roman*)}

\newcolumntype{L}[1]{>{\hsize=#1\hsize\raggedright\arraybackslash}X}%
\newcolumntype{R}[1]{>{\hsize=#1\hsize\raggedleft\arraybackslash}X}%
\newcolumntype{C}[1]{>{\hsize=#1\hsize\centering\arraybackslash}X}%
\newcolumntype{Y}{>{\centering\arraybackslash}X}%

\Title{Performance and Energy Trade-Offs for Parallel Applications on Heterogeneous Multi-Processing Systems}

\Author{Demetrios~A.~M.~Coutinho $^{1,}$*, Daniele~De~Sensi $^{2}$, Arthur~Francisco~Lorenzon $^{3}$, Kyriakos~Georgiou $^{4}$, Jose~Nunez-Yanez $^{4}$, Kerstin~Eder $^{4}$, Samuel~Xavier-de-Souza $^{5}$}
\AuthorNames{Demetrios~A.~M.~Coutinho, Daniele~De~Sensi, Arthur~Francisco~Lorenzon, Kyriakos~Georgiou, Jose~Nunez-Yanez, Kerstin~Eder, Samuel~Xavier-de-Souza}

\address{%
$^{1}$ \quad Instituto Federal do Rio Grande do Norte, Pau dos Ferros 59900-000, Brazil;\\
$^{2}$ \quad Computer Science Department, Università di Pisa, Pisa 56127, Italy; desensi@di.unipi.it (D.D.S.)\\
$^{3}$ \quad Computer Science Department, Universidade Federal de Pampa, Alegrete 97546-550, Brazil; aflorenzon@unipampa.edu.br (A.F.L.)\\
$^{4}$ \quad Department of Computer Science, University of Bristol, Bristol BS8 1UB, UK; kyriakos.georgiou@bristol.ac.uk (K.G.); j.l.nunez-yanez@bristol.ac.uk (J.N.-Y.); kerstin.eder@bristol.ac.uk (K.E.)\\
$^{5}$ \quad Department of Computer Engineering and Automation, Universidade Federal do Rio Grande do Norte, Natal 59078-970, Brazil; samuel@dca.ufrn.br (S.X.d.S.)
}
\corres{Correspondence: demetrios.coutinho@ifrn.edu.br (D.A.M.C.)}

\abstract{This work proposes a methodology to find performance and energy trade-offs for parallel applications running on Heterogeneous Multi-Processing systems with a single instruction-set architecture. These offer flexibility in the form of different core types and voltage and frequency~pairings, defining a vast design space to explore. Therefore, for a given application, choosing a configuration that optimizes the performance and energy consumption is not straightforward. Our~method proposes novel analytical models for performance and power consumption whose parameters can be fitted using only a few strategically sampled offline measurements. These models are then used to estimate an application's performance and energy consumption for the whole configuration~space. In turn, these offline predictions define the choice of estimated Pareto-optimal configurations of the~model, which are used to inform the selection of the configuration that the application should be executed on. The methodology was validated on an ODROID-XU3 board for eight programs from the PARSEC Benchmark, Phoronix Test Suite and Rodinia applications. The generated Pareto-optimal configuration space represented a 99\% reduction of the universe of all available configurations. Energy savings of up to 59.77\%, 61.38\% and 17.7\% were observed when compared to the \texttt{performance}, \texttt{ondemand} and \texttt{powersave} Linux governors, respectively, with higher or similar performance.}


\begin{document}

\section{Introduction}

Low energy consumption is a key requirement in the design of modern embedded systems, affecting the size, cost, user~experience, and the capability to integrate more high-level features.
Single Instruction-Set Architecture (Single-ISA) Heterogeneous Multi-Processors (HMP) are known for delivering a significantly higher performance-power ratio than their counterparts. There exist many commercially available designs in the embedded and mobile world~\cite{pharm06overview,bigLITTLE,chitlur2012quickia}. Nevertheless, in this type of architecture it is increasingly more complicated to find the number of cores, operating frequency, and voltage that optimize performance and energy consumption to meet the requirements of a given application~\cite{Wang2015,koufaty2010bias,saez2017towards}.
This complexity increases when application characteristics need to be extracted at runtime to increase performance or save energy at different phases during the execution of an application~\cite{sawalha2011}.

One of the commercial widely used heterogeneous architectures is the ARM big.LITTLE~\cite{bigLITTLE}. It~consists of two clusters with two types of processing cores, each one containing one or more cores. Cluster type \emph{big} is composed of higher performance larger cores that are also more power-hungry. Cluster type \emph{LITTLE} consists of smaller cores that are slower and more energy-efficient. The increased sophistication of this type of architecture delivers a challenging task to develop better energy management solutions. Running a multi-threaded application only on big cores may not justify the performance gain related to energy consumption. Also, using only small cores may not be the best choice for reducing energy consumption, as the application may suffer a significant increase in execution time. Besides, it is possible to scale up or scale down the core frequencies to improve the performance or save energy, respectively. Each arrangement, or configuration, of operating frequency and the number of available cores of each cluster type to execute a given application may produce a different performance-energy trade-off.

As simple approach to obtaining one or more configurations in HMPs that provide most beneficial performance-energy trade-offs is to collect the performance and energy consumption of a given application in all available configurations. However, in current heterogeneous systems, this would require a high cost to be performed on-line. For example, an embedded system with two clusters each with four cores and 16 clock frequency levels yields 4096 configurations to explore. Such a large number of configurations could be unfeasible for on-line approaches since, in many cases, the~search process for the best configuration can outlast the execution itself. Even if executed offline, although it could be possible to evaluate all configuration, the expected overheads will make the method unfeasible.

Considering the vast diversity of performance-energy trade-offs that the large configuration space of HMP systems yields for a given application, the understanding of these trade-offs becomes critical to energy-efficient software development and operation~\cite{Jin2017}. We present a novel methodology to find the best performance and energy trade-off configurations of parallel applications running on (HMP) systems. It is intended to be used by the operating system to make scheduling decisions at runtime according to performance and energy consumption requirements of the given application and according to the system it is running on. For that, we propose analytical models for performance and power that are fitted offline using a few measurements of the application's execution.

The combination of the outcomes of the performance and power models yields the energy model that can be used to estimate the energy consumption of all configurations for a given application on a specific architecture. The performance and energy models are used to assess the whole configuration space, commonly in the order of thousands of points, and the outcomes are used to select the configurations that lay on the performance and energy Pareto frontier, whose number is often in the order of tens of points. The Pareto frontier is the set of all optimal solutions from which it is not possible to improve a criterion (performance or energy) without degrading another. Hence, we propose a methodology that can estimate the best relationship between performance and energy consumption, without requiring an expensive exploration of the full search space.

We employed the methodology in an ODROID-XU3 board on eight multi-thread applications.  We~present the fitting of the proposed models and the trade-off Pareto frontier configurations for each~application. The results show that the proposed methodology and models can successfully estimate the Pareto trade-off consistently with measured data. On average, we obtained gains in performance and energy consumption in the order of 33\% and 40\% respectively, when compared to the \texttt{performance}, \texttt{ondemand} and \texttt{powersave} standard Linux DVFS governors. 
The main contributions of this paper are:
\begin{itemize}
    \item Simple analytical models for HMP performance and power consumption that only require measurements of execution time and power, avoiding performance hardware counters, which may not be accessible on all architectures.
    \item  A practical methodology to sample the configuration space of a given application efficiently and select points that are estimated to be Pareto-optimal regarding trade-offs of performance and energy consumption.
    \item Results show that the specific knowledge of the application performance and the system power embedded into analytical models can deliver significantly better results than conventional DVFS~governors.
    \item The technique is easy to use on new applications running on the same HMP system; the power model of the HMP system can be reused, and the performance model can be re-fitted based solely on the execution time of the new application.

\end{itemize}

The rest of this paper is arranged as follows. Section~\ref{sec:related} presents our motivation and places this paper in the context of existing research. Section~\ref{sec:methodology} introduces our modeling techniques and our methodology to find the trade-off configurations. Section~\ref{sec:Experimental} describes the experimental setup used to evaluate our methodology, and compares our results to those obtained using the standard Linux DVFS governors. Our conclusions and a discussion of future work follow in Section~\ref{sec:conclusions}.

\section{Related Work}
\label{sec:related}

We attempt to distinguish the main characteristics that differentiate our work from others with respect to existing approaches for finding applications' configurations to improve energy-efficiency. The most relevant works are characterised in Table~\ref{tab:relatedwork}.
\begin{table}[H]
    \centering
    
    \caption{Related work categorized by various aspects.}
        \label{tab:relatedwork}
    \scalebox{.95}[1.0]{\resizebox{\textwidth}{!}{%
    \begin{tabular}{lccccccccccccc}
            \noalign{\hrule height 1.0pt} 
            & & & & & & & & & & & & & \\ [-2ex]
            \multicolumn{1}{c}{} & \multicolumn{5}{c}{\textbf{Scenarios}} & \multicolumn{3}{c}{\textbf{Techniques}} & \multicolumn{3}{c}{\textbf{Decision guided by}} & \multirow{2}{*}[.2cm]{\textbf{Decision Strategy}}  & \multicolumn{1}{c}{\textbf{Arch.}} \\ 
             \cline{1-14} 
             & & & & & & & & & & & & & \\ [-2ex]
            \multicolumn{1}{c}{\textbf{Reference}} & \textbf{\rotatebox[origin=c]{90}{Multi-threaded}} & \textbf{\rotatebox[origin=c]{90}{Multi-apps}} & \textbf{\rotatebox[origin=c]{90}{Power}} & \textbf{\rotatebox[origin=c]{90}{Energy}} & \textbf{\rotatebox[origin=c]{90}{Performance}} & \textbf{\rotatebox[origin=c]{90}{DVFS}} & \textbf{\rotatebox[origin=c]{90}{DPM}} & \textbf{\rotatebox[origin=c]{90}{Placement}} & \textbf{\rotatebox[origin=c]{90}{Monitoring}} & \textbf{\rotatebox[origin=c]{90}{Profiling}} & \textbf{\rotatebox[origin=c]{90}{Prediction}} & \textbf{\rotatebox[origin=c]{90}{Pareto}}  & 
            \textbf{\rotatebox[origin=c]{90}{Heterogeneous}}
            \\ \hline
            & & & & & & & & & & & & & \\ [-2ex]
            \multicolumn{1}{l}{Endrei et. al~\cite{Endrei2018}} & X & - & - & X & X & X & - & - & - & - & Analytical & X & -  \\ \hline
            & & & & & & & & & & & & & \\ [-2ex]
            \multicolumn{1}{l}{De Sensi et al. ~\cite{taco16}} & X & - & X & - & X & X & - & X & X & - & ML & -  & - \\ \hline
            & & & & & & & & & & & & & \\ [-2ex]
            \multicolumn{1}{l}{Gupta et al.~\cite{Gupta2017}} & X & X & - & X & X & X & X & - & X & X & - & X  & X \\ \hline
            & & & & & & & & & & & & & \\ [-2ex]
            \multicolumn{1}{l}{Manumachu et al.~\cite{Manumachu2018}} & X & - & - & X & X & X & - & - & - & - & Analytical & X & - \\ \hline
            & & & & & & & & & & & & & \\ [-2ex]
            \multicolumn{1}{l}{Tzilis et al. ~\cite{Tzilis2019}} &  -  & X & - & - & X & X & X & X & X & X & Analytical &- & X \\ \hline
            & & & & & & & & & & & & & \\ [-2ex]
            \multicolumn{1}{l}{Loghin et al.~\cite{Loghin2018}} & X & - & - & - & X & - & - & - & - & - & Analytical & -  & X \\ \hline
            & & & & & & & & & & & & & \\ [-2ex]
            \multicolumn{1}{l}{ Vasilakis et al.~\cite{Vasilakis2017}} & X & - & - & X & X & X & - & X & - & - & Analytical & -  & X \\ \hline
            & & & & & & & & & & & & & \\ [-2ex]
            \multicolumn{1}{l}{De Sensi~\cite{DeSensi2016}} & X & - & X & - & X & X & - & - & - & - & Analytical & - & - \\ \hline
            & & & & & & & & & & & & & \\ [-2ex]
            \multicolumn{1}{l}{Aalsaud et al.~\cite{Aalsaud2016}} & - & X & X & - & - & X & X & X & - & X & - & -  & X \\ \hline
            & & & & & & & & & & & & & \\ [-2ex]
            \multicolumn{1}{l}{\textbf{Proposed Work}} & \textbf{X} & \textbf{-} & \textbf{-} & \textbf{X} &  \textbf{X} & \textbf{X} & \textbf{-} & \textbf{X} & \textbf{-} & \textbf{X} & \textbf{Analytical} & \textbf{X}  & \textbf{X} \\ \noalign{\hrule height 1.0pt} 
        \end{tabular}%
    }}
        
\end{table}

Research can be categorized concerning the application scenario focus. Some works concentrate on concurrent  execution of multiple applications~\cite{Tzilis2019,Gupta2017,Aalsaud2016} and others exclusively on single-application scenarios~\cite{Endrei2018,taco16,Manumachu2018}. Tzilis  et al.~\cite{Tzilis2019} propose a runtime manager that estimates the performance and power of applications choosing the most efficient configuration by making use of heuristics to select candidate solutions. They do not consider multi-threaded versions, but instead, they allow multiple instances of the same application to run simultaneously. Indeed, energy efficient management of concurrent applications is harder to accomplish due to dynamically changing situations. However, guarantee an energy consumption requirement with a minimum performance level of a single application affects the energy efficiency of a whole cluster-based system. 

Thus, considering this aspect important, our methodology manages energy and performance requirements of individual applications as approached by other works~\cite{Endrei2018,Gupta2017,Manumachu2018,Vasilakis2017}. Some works are only concerned about the performance constraints~\cite{Tzilis2019,Loghin2018}.  Furthermore, for parallel or multi-threaded applications, some works like ours extend Gustafson's~\cite{gustafson88reevaluating} and Amdahl's law~\cite{amdahl67validity} in order to characterize the application~\cite{Loghin2018,DeSensi2016,Woo2008}. Others attempt to identify the phases of each application thread through monitoring performance counters~\cite{Gupta2017,sawalha2011}. As shown in Loghin et al.~\cite{Loghin2018}, the applications' workloads with large sequential fractions present small energy savings regardless of the heterogeneous processing system. Therefore, it is clearly vital to exploit the energy efficiency in HPM architectures as well as the parallelism of single multi-threaded applications. 

%
A notable number of power-management approaches target a reduction of power dissipation and a performance increase by combining three techniques. First, Dynamic Voltage and Frequency Scaling (DVFS)~\cite{Herbert2007}, which selects the optimal operating frequency. Second, Dynamic Power Management (DPM)~\cite{Benini2000}, which turns off system components that are idle. Finally, application placement (allocation)~\cite{Singh2013}, which determines the number of cores or the cluster type an application executes on.
%
%
Some strategies decide the application placement without taking control of the DVFS aspect~\cite{Loghin2018, Donyanavard2016, Sarma2015, Kim2014}. 
Gupta et al.~\cite{Gupta2017} and Tzilis et al.~\cite{Tzilis2019} have different approaches, yet combine DVFS, DPM, and application placement by setting the operating frequency/voltage and the type and number of active cores simultaneously. Also, there are works where the only concern is finding the best cluster frequency performances~\cite{Endrei2018,Manumachu2018,Loghin2018,DeSensi2016}.
In this work, we find the Pareto-optimal core and frequency configurations in heterogeneous architecture that deliver the optimal performance-energy trade-off. Consequently, our method merges DVFS and DPM  by choosing the operating frequency/voltage and number of active cores of each cluster type. Thus, our approach covers two of the main techniques for saving energy.
Next, we describe the points that are essential to guide the configuration choice. 
Generally, we distinguish:
\begin{mylist}[label=(\roman*)]
  \item offline application profiling;
  \item runtime performance or power monitoring;
  \item predictions performed using a model. 
\end{mylist} 
\mbox{Gupta et al.~\cite{Gupta2017}} characterize the application by collecting power consumption, processing time and six performance counters. They use the data to train classifiers that map performance counters to optimal configurations.  Then, at runtime, these classifiers and performance counters are used to select the optimal configuration concerning a specific metric, e.g. energy consumption.
The major problem of this approach is the growth of the number of cores and cluster types, leading to this strategy losing reliability. \mbox{Tzilis et al.~\cite{Tzilis2019}} use a strategy that requires a total number of runs that is linear concerning the number of cores. They use a similar number of performance counters as~\cite{Gupta2017} to profile the application. When the profiled application spawns, they match the online measurement to the closest profiled value and use it as a starting point to predict its performance in the current situation. De~Sensi et al.~\cite{taco16} focus on a single application and also monitor data to refine power consumption and throughput prediction~models. They compared the predicted outcomes from their models against the data monitored in the current configuration. In case the prediction error is lower than a defined threshold, the calibration phase completes. The problems of online monitoring are the overhead incurred in refining models or making decisions.  In contrast, our proposal characterises the application performance and power consumption of a given HMP system using only a few strategically sampled offline measurements. Further, they are used for predicting the optimal performance and energy consumption trade-off configurations.  

A sort of performance and power modelling is commonly used to guide the configuration selection. Many works~\cite{taco16,Gupta2017, Tzilis2019,Vasilakis2017, Aalsaud2016} are based on analytical models that rely upon runtime measurements retrieved from traces or instrumentation to collect performance counters, which is often not a trivial process. Usually, performance and energy events should be recorded in different runs to prevent counter multiplexing. Otherwise, it may cause application disturbance and decrease measurement accuracy~\cite{Endreibook}.  As in other works~\cite{Endrei2018, Manumachu2018, DeSensi2016}, our approach only relies on a minimum set of parameters, such as the number of cores of each cluster and the cluster frequency. Thus, our approach is simple to automate because it does not require any external instrumentation tools, and its utilisation across different architectures is less restricted. 

Loghin and Teo~\cite{Loghin2018} derived equations to evaluate the speedup and energy savings of modern shared-memory homogeneous and heterogeneous computer systems. They introduced two parameters: the inter-core speedup (ICS), describing the execution time relation among diverse types of cores in a heterogeneous system for a given workload; and the active power fraction (APF), representing the ratio between the average active power of one core and the idle power of the whole system. They validated their work against measurements on different types of heterogeneous and homogeneous modern multicore~systems. Their results show that energy savings are limited by the system's APF, particularly on performance cores and by the workload's sequential fraction when running on more efficient cores. 
We~used a similar approach when devising our performance and power equations. Our performance model has a parameter that represents the median of the larger (big) cores speedup compared with the smaller (LITTLE) cores. The value of this parameter can be determined independently of any target application, i.e., it is application-agnostic.  Our power model describes the dynamic and the static power consumption of the active cores of each cluster~type.  
The aim is to capture the properties of the hardware architecture so that we do not need to fit the power model and the performance speedup parameter for each application. Ultimately, our energy model is a combination of a performance model and an application-agnostic power model. 
The main concern with using heuristics, models refinement or any other strategy to make decisions at runtime is the added overhead. Thus, the most significant contribution of our work is to avoid a full exploration of the search space to find the best configurations to execute an application at~runtime. A similar goal, but with a different approach, has been pursued by De Sensi~\cite{DeSensi2016}.  However, the architecture is not HMP, and it offers only 312 possibles configurations. Moreover, \mbox{Endrei et. al~\cite{Endrei2018} } and Manumachu et al.~\cite{Manumachu2018} use Pareto Frontier as a strategy to minimize energy consumption without degrading application performance. Still, they do not work on ISA heterogeneous architecture as \mbox{Gupta et al.~\cite{Gupta2017}} and our approach. Furthermore, our approach has the advantage of not depending on performance counters, resulting in less restricted employment. 

Using the Pareto frontier method, we can find optimal configurations, since it represents the optimal trade-off between energy consumption and performance for the target system~\cite{Balaprakash2014}. Thus, users can take advantage of configurations at the Pareto frontier to guide development toward energy-optimal computing~\cite{Sen2017}. Also, these configurations can provide a power-constrained performance optimization to identify the optimized performance under a power budget~\cite{Bailey2015}. Moreover, the Pareto-optimal fronts offer a trade-off zone that can be used to produce the optimal performance and energy efficiency prediction of a model~\cite{Endrei2018,vsli_demetrios}.

In summary, we propose a method to find the optimal energy-performance trade-offs on a two-cluster heterogeneous architecture of a single multi-threaded application. The energy estimation is obtained by combining the offline performance and power models. The offline performance application characterization is straightforward; it does not use performance counters. The power model involves only architecture-specific parameters, so it is sufficient to fit only once for a specific platform. The individual applications’ performance and power requirements can be achieved by the energy-performance trade-off zone provided by the Pareto frontier.

\section{Estimating Pareto-Optimal HMP Configurations}
\label{sec:methodology}
\begin{figure}[H]
  \centering
  \includegraphics[scale=0.4]{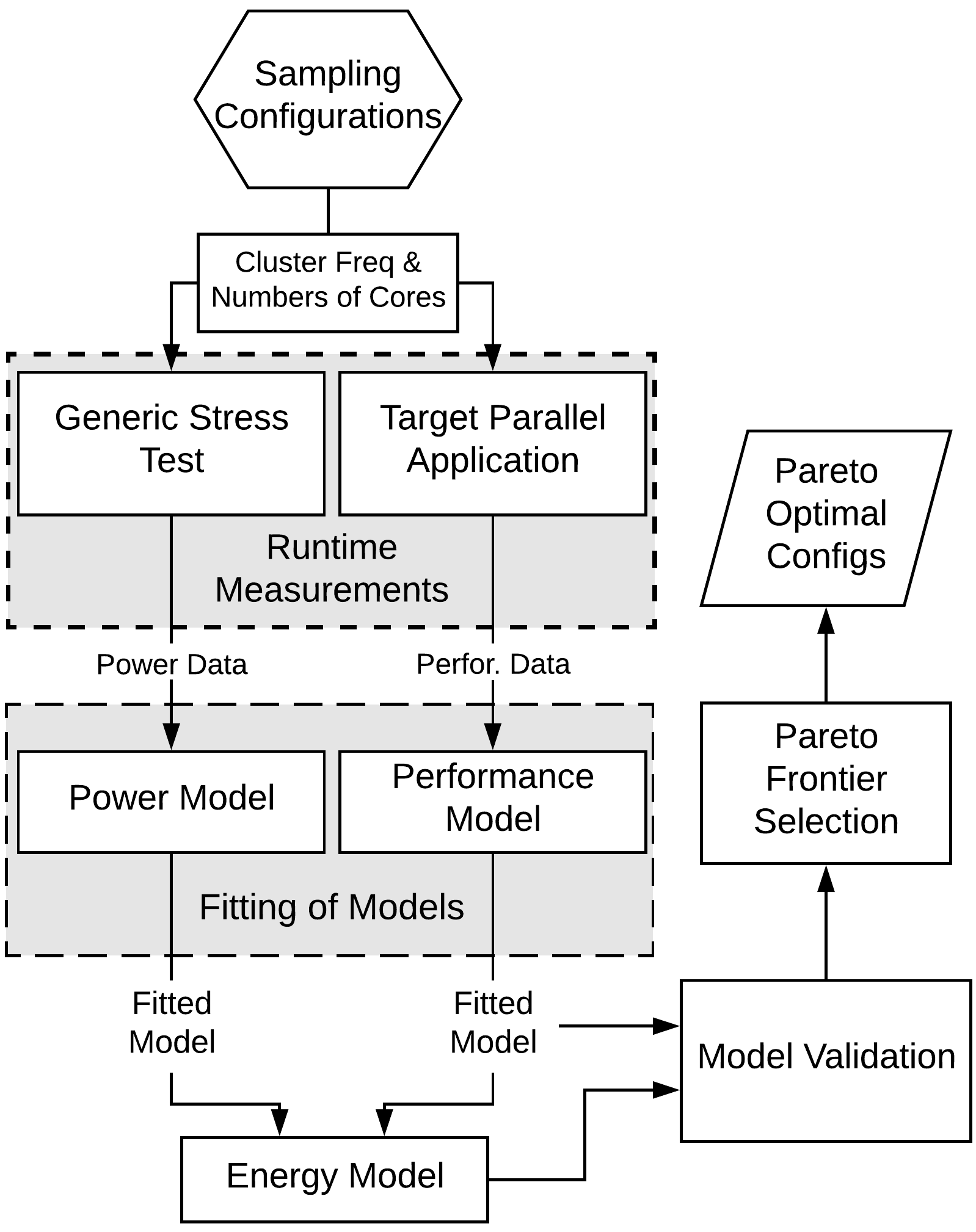}
  \caption{Flowchart of the methodology applied for predicting the optimal performance and energy efficiency trade-offs of a parallel application.}
  \label{fig:fittingscheme}
\end{figure}

This section describes the methodology proposed to estimate the Pareto-optimal configurations for HMP systems. A configuration is defined as a point in the vast parametric space that defines parallel software operation, including the number of processing cores used in each cluster and the operating frequency of those cores. We assume that the clusters have their own voltage and frequency~domain, which is shared among the cores of the cluster. Design attributes such as the number of clusters and the number of cores and frequency levels of each cluster are known parameters. The model matches a large subset of actual HMP systems, to which the proposed methodology can be applied. Five steps compose this methodology: sampling of configurations, runtime measurements, fitting of models, model validation, and the Pareto frontier selection. Figure~\ref{fig:fittingscheme} presents an overview of these steps. 

In the first step---sampling of configurations, we need to carefully choose the hardware configurations that will be used to obtain the required data to fit the parameters of the models. To~obtain parameters that generate higher accuracy, data measurements should be as diverse as possible, i.e., two or more measurement points should not have a similar number of cores and operating frequency, since~those points will most likely not add significant information to the modeling. For this reason, we~used a Halton low discrepancy number generator to choose configurations that are not too similar to each other. Morokoff et al.~\cite{MorokoffC94} describe how the low discrepancy number generator ensures that it covers the domain more evenly compared to pseudo-random number generators. Also, De~Sensi~\cite{DeSensi2016} shows that by picking equidistributed points, it is possible to achieve higher accuracy compared to pseudo-random selection of the points. 

In the second step---runtime measurements, we evaluate the sampled configurations at runtime, measuring performance and power data to fit the models. First, for each configuration, we execute a generic stress-test to stress the processor and capture power consumption. Second, we execute the target parallel application to measure execution time for another subset of sampled configurations using the Halton number generator. 

In the third step---fitting of models, we fit the measured data to the proposed architecture- and application-specific performance model and the architecture-specific, application-agnostic power~model. 
The performance model intends to estimate the execution time of a parallel application for any point in the configuration space. We adjust the parameters of the power and performance models separately. The goal of the proposed power model is not to be accurate per se, but to capture the trend changes in power when a change of the operating frequency and number of active cores occurs, independent of the application running. The power consumption prediction accuracy is not strictly crucial if there are no power restrictions to apply on system budget. More importantly, we aim for decreasing energy consumption in HMP since it has more impact on batteries, which is essential on embedded devices. Moreover, being application-agnostic avoids the need for power measurements for every application while preserving the architecture's signature. Thus, we do not need to fit the power model for each application. Besides, the process of acquiring these measurements also increases the difficulty of automating these models as practical tools.
For each model, we use a non-linear regression minimizing the Root Mean Square Error, given by
\begin{equation}
     \text{RMSE} = \sqrt{\frac{\sum_{i=1}^{n}(A_i - E_i)^2}{n}},
\end{equation}
where \(A_i\) is the \(i\)th measured data of \(E_i\), \(E_i\) is the \(i\)th estimated value given by the model, and  \(n\) is the total number of configurations. 
Sections~\ref{sec:performance} and~\ref{sec:power} detail the proposed performance and power consumption models, respectively. 
The proposed energy model is a combination of the power consumption model---previously fitted to a specific HMP system, and the performance model---fitted to a given parallel application and the same HMP system. Section~\ref{sec:energy} presents the proposed energy~model.

In the fourth step---model validation, we evaluate the performance and energy consumption models for every point in the vast configuration space that defines parallel software operation. 
 
Finally, in the fifth step---Pareto frontier selection, we select all modeled configurations whose models' outcomes lay on the Pareto frontier, i.e., those configurations that yield the optimal performance and energy consumption trade-offs.

\subsection{Application Performance Modelling}
\label{sec:performance}
In this section, we devise a performance model for an HMP platform with two processing clusters. We assume that a parallel application coherently runs in $b$ big cores and $L$ LITTLE cores. Moreover, we do not consider or require different frequencies for each core in a cluster, i.e., all the cores in a cluster run at the same frequency. Inspired by other works~\cite{hill09amdahl,barros2015optimal,morad2006performance}, we devised the following model that can be used to estimate the performance of a given parallel application running on a two-cluster HMP.
\begin{equation}
T_{\rm HMP}(F, F_{\rm b},F_{\rm L},b,L) = T_{\rm L}(F)\left(\frac{(1-f)F}{{\rm perf}\cdot~F_{\rm b}} + \frac{fF}{b\cdot{\rm perf}\cdot~F_{\rm b} + LF_{\rm L}}  \right)\text{, if } {\rm b} > 0, 
\label{eq:performance}
\end{equation}
where $T_{\rm HMP}$ is the performance target, that is the total execution time goal for a given application. $T_{\rm L}(F)$ is the application's execution time running on a single LITTLE core with operating frequency $F$, which is not necessarily the same frequency of $F_{ \rm L}$.  
The number of active big and LITTLE cores in the processor is denoted by $b$ and $L$, respectively.
$F_{\rm b}$ and $F_{\rm L}$ are the operating frequencies of the big and LITTLE cores, respectively. \textit{\texttt{\emph{perf}}} is the performance improvement when moving computation from a LITTLE to a big core, independent of an application. Note that, \textit{\texttt{\emph{perf}}} depends on the hardware design, so we assume it has a fixed value representing how fast a big core is when compared to any LITTLE core. Thus, we do not need to fit this performance speedup parameter for each application.

The value of $f$ represents the parallel fraction of the application. It is the only parameter that characterizes the application. Indeed, by modelling complex parallel applications using their parallel fraction only, we are neglecting the effects that a frequency change has on the performance of the memory hierarchy, on the parallel overhead, and on the distribution of load across the heterogeneous cores~\cite{furtunato2019parallel}. Besides, as those features should limit the parallel speedup, it is expected that the sequential fraction includes them. It is our priority to keep our model straightforward to achieve low runtime overheads. We observed that, despite the many simplifications, the model still provides reasonably accurate energy consumption estimations.

We consider that the sequential fraction of the application is accelerated by using one of the big cores when such core is available. In this approach, the sequential portion expects to include the parallel overhead, i.e., communication or synchronization among the threads, which does not accelerate by making the big core run faster---as it would do for actual sequential code. Otherwise, we~can extend Equation~\eqref{eq:performance} as follows:
\begin{equation}
\begin{split}
T_{\rm HMP}(F, F_{\rm b},F_{\rm L},b,L) = T_{\rm L}(F)\left(\frac{(1-f)F}{F_{\rm L}} + \frac{fF}{LF_{\rm L}}\right)\text{, if } {\rm b} = 0.
\label{eq:performance2}
\end{split}
\end{equation}

In Equation~\eqref{eq:performance2}, when there is no big core available or active, the application would not take advantage of the performance improvements from the big cores, therefore the parameter \textit{\texttt{\emph{perf}}} is~removed. Also, the big core operating frequency is not used by the parallel application. 

This performance model assumes that the modelled parallel workload is dynamically distributed over the running threads. If that is not the case for the application, cores may run out of work and become idle while others might become overloaded and delay the end of the program's execution. Thus, the performance model may then make incorrect estimations as idle cores still consume power. This can lead to improper power predictions, resulting in inadequate energy predictions. Although, this is a limitation of the proposed approach, the advantages in terms of performance gain and energy reduction presented in this work is an argument for pursuing this type of workload balancing scheme.

\subsection{Power Consumption Modelling}
\label{sec:power}
In this section, we devise a power model for an HMP platform with two clusters. 
Assuming that the transistors used on both types of cores have the same power consumption behaviour with respect to their operating frequency and voltages, we can model the power consumption of the big and LITTLE cores, as follows:
\begin{equation}
P_{\rm HMP}(F_{\rm b},F_{\rm L},b,L) = P_{\rm b}(F_{\rm b},{ b}) + P_{\rm L}(F_{\rm L},{ L}),
\label{eq:power1}
\end{equation}
where $P_{\rm b}(F_{\rm b},{ b})$
is the power consumption of $b$ big cores running at frequency $F_{\rm b}$, $P_{\rm L}(F_{ \rm L},{ L})$ is the power consumption of the $L$ LITTLE cores running at frequency $F_{\rm L}$. 

Consider the following equation to describe the power consumption of a processor running at frequency $F$~\cite{Brodersen1995,Alonso2014,Kim2003}.
\begin{equation}
P(F) = aCV^2F + VI_{\rm Leak},
\label{eq:powercore1}
\end{equation}
where $a$ is the average activity factor, $C$ is the load capacitance, $V$ is the supply voltage, and $I_{\rm Leak}$ is the average leakage current. Also, consider the equation that defines the maximum operating frequency for a transistor~\cite{Suleiman2005,Venkatachalam2005}:
\begin{equation}
F_{\rm Max} = \kappa_1\frac{(V-V_{\rm Th})^h}{V},
\label{eq:freqmax}
\end{equation}
where $V_{\rm Th}$ denotes the threshold voltage, $\kappa_1$ is a constant, and $h$ is a  technology-dependent value often assumed to be within the range \texttt{[2,3]}. 
Since performance is proportional to the operating~frequency, digital circuits often operate with voltage and frequency pairs that push frequency closer to the~maximum.
If maximum performance is not required, the voltage is scaled down with frequency to maintain this policy~\cite{CARDOSO201717}. 
Usually, device vendors provide a table of discrete values of supply voltages that the processing chip can operate beneath by maximum frequencies ~\cite{Tang2017}. 
Generally, the supply voltage $V$ can be perceived as a linear function of the frequency depicted such as follows:
\begin{equation}
    V = \kappa_2 F\text{, }
    \label{eq:freq_volt}
\end{equation}
where $\kappa_2$ is any constant. Some authors make a similar approach~\cite{CARDOSO201717, Usman2013} showing that it is reasonable and common in the literature to make such approximation. 
Furthermore, we reiterate that we intend to capture the trend of how a change in the operating frequency-voltage would affect the power consumption of the system. For this work, our target is to save energy consumption and not accuracy in power consumption prediction. Using Equation~\eqref{eq:freq_volt}, we can rewrite Equation~\eqref{eq:powercore1} as follows:
\begin{equation}
\begin{split}
P(F) &= aC(\kappa_2 F)^2F + \kappa_2 FI_{\rm Leak},\\
&= aC\kappa_2^2 F^3 + \kappa_2 FI_{\rm Leak}.\\
\end{split}
\label{eq:powercore2}
\end{equation}

Considering $I_{\rm Leak}$ and $a$ to be approximately constant, we can approximate Equation~\eqref{eq:powercore2} by
\begin{equation}
P(F) = \alpha F^3 +\beta F,
\label{eq:powercore3}
\end{equation}
where
$\alpha = aC\kappa_2^2$ 
and     
$\beta = \kappa_2 I_{\rm Leak}$
are considered constants that mainly abstract semiconductor technology attributes. These attributes should be application-independent and fixed since they rely on hardware design. Then, the power of each cluster type is modelled as follows:
\begin{equation}
P_{\rm b}(F_{\rm b},{ b}) = { b} \alpha_{\rm b} F_{\rm b}^3 + C_{\rm b}\beta_{\rm b} F_{\rm b},
\label{eq:powerb}
\end{equation}
\begin{equation}
P_{\rm L}(F_{\rm L},{ L}) = { L} \alpha_{\rm L} F_{\rm L}^3 + C_{\rm L}\beta_{\rm L} F_{\rm L},
\label{eq:powerL}
\end{equation}
where ${ b}$ and ${ L}$ are the numbers of active big and LITTLE cores, respectively, accounted for the dynamic power modelling component. The technical setup of our experimental platform did not permit disabling individual cores. Therefore, we consider a single term of leak current for each cluster. Then, $ C_{\rm b}$ and $ C_{\rm L}$ represent the total number of cores of each cluster type and account for the static power modelling component.

Note that this power model does not distinguish one application from another since the activity factor $a$ is assumed to be constant. This is a very generic assumption, which may cause an inaccurate power consumption estimation. Nevertheless, for this approach, more critical than accurate estimations of power is to capture how a change of the operating frequency-voltage would affect power consumption. It is expected that the way in which this change affects the power is the~same, regardless of the activity factor. In other words, as long as the activity factor of the target application does not change too much from one frequency configuration to another, this assumption will not adversely impact on the proposed power model.


By combining Equations~\eqref{eq:power1},~\eqref{eq:powerb} and~\eqref{eq:powerL}, we come to the power model for the whole HMP chip parameterized by the two frequencies of each cluster as follows:
\begin{equation}
  P_{\rm HMP}(F_{\rm b},F_{\rm L},b,L) = P_{{\rm HMP_P}} = b\alpha_{\rm b}F_{\rm b}^3 + C_{\rm b}\beta_{\rm b}F_{\rm b} + L\alpha_{\rm L}F_{\rm L}^3 + C_{\rm L}\beta_{\rm L}F_{\rm L}.
\label{eq:powerHMP}
\end{equation}

Equation~\eqref{eq:powerHMP} represents the power consumption of a parallel fraction ($P_{{\rm HMP_P}}$) of an parallel application. Also, we need to predict the power of the sequential part of a code that runs in one core. 
\begin{equation}
    P_{{\rm HMP_S}}(F_{\rm b},F_{\rm L}) =\\
  \begin{cases}
  \alpha_{\rm b}F_{\rm b}^3 + C_{\rm b}\beta_{\rm b}F_{\rm b} + C_{\rm L}\beta_{\rm L}F_{\rm L},  & \; \text{if } b > 0, \\
   C_{\rm b}\beta_{\rm b}F_{\rm b} + \alpha_{\rm L}F_{\rm L}^3 + C_{\rm L}\beta_{\rm L}F_{\rm L}, & \; \text{otherwise.}
  \end{cases}
\label{eq:powerSequential}  
\end{equation}

When there is, at least, one big core available, we consider the dynamic power of one big core and the static power of all cores. In the case that just LITTLE cores are available, then the sequential part runs on one LITTLE core. In this case, the power consumption of the sequential portion contains the static power of all cores and the dynamic component of one LITTLE core.

\subsection{Energy Modelling}
\label{sec:energy}

In this section, we devise an energy model for a parallel application running on a two-cluster HMP platform by aggregating the sequential and parallel energy consumption as follows:
\begin{equation}
E_{\rm HMP}(F, F_{\rm b},F_{\rm L},b,L)= E_{{\rm HMP_S}} + E_{{\rm HMP_P}}.
\label{eq:energy}
\end{equation}

By combining the parallel part of Equation~\eqref{eq:performance} and the power model of the whole chip in Equation~\eqref{eq:powerHMP}, we have the parallel portion energy model component:
\begin{equation}
E_{{\rm HMP_P}} =  \frac{T_{\rm L}(F)fF\left({L\alpha_{\rm L}F_{\rm L}^3 + C_{\rm L}\beta_{\rm L}F_{\rm L}} + {b\alpha_{\rm b}F_{\rm b}^3 + C_{\rm b}\beta_{\rm b}F_{\rm b}}\right)}{b\cdot{\rm perf}\cdot~F_{\rm b} + LF_{\rm L}}.
\label{eq:energyParallel}
\end{equation}

Using Equation~\eqref{eq:powerSequential} and the sequential part of the Equations~\eqref{eq:performance} and~\eqref{eq:performance2}, we have the sequential portion energy model component, which relies on the number of active big cores available:
\begin{equation}
    E_{{\rm HMP}_S} =
  \begin{cases}
    \displaystyle
   \frac{T_{\rm L}(F)(1-f)F\left({C_{\rm L}\beta_{\rm L}F_{\rm L}} + {\alpha_{\rm b}F_{\rm b}^3 + C_{\rm b}\beta_{\rm b}F_{\rm b}}\right)}{{\rm perf}\cdot~F_{\rm b}}, & \; \text{if }  b > 0, \\[0.2cm]
    \displaystyle
   \frac{T_{\rm L}(F)(1-f)F\left(C_{\rm b}\beta_{\rm b}F_{\rm b} + \alpha_{\rm L}F_{\rm L}^3 + C_{\rm L}\beta_{\rm L}F_{\rm L}\right)}{F_{\rm L}}, & \; \text{otherwise.}
  \end{cases}
\label{eq:energySequential}  
\end{equation}

Finally, combining Equations~\eqref{eq:energyParallel} and~\eqref{eq:energySequential}, we come to the consolidated energy model for the whole HMP chip parameterized by the frequencies and the number of active cores in each cluster:
\begin{equation}
\begin{split}
&E_{\rm HMP}(F, F_{\rm b},F_{\rm L},b,L) = T_{\rm L}(F)F~\times \\
& \begin{cases}
\displaystyle
\frac{(1-f)(C_{\rm L}\beta_{\rm L}F_{\rm L} + \alpha_{\rm b}F_{\rm b}^3 + C_{\rm b}\beta_{\rm b}F_{\rm b})}{{\rm perf}\cdot~F_{\rm b}} + 
\frac{f(L\alpha_{\rm L}F_{\rm L}^3 + C_{\rm L}\beta_{\rm L}F_{\rm L} + b\alpha_{\rm b}F_{\rm b}^3 + C_{\rm b}\beta_{\rm b}F_{\rm b})}{b\cdot{\rm perf}\cdot~F_{\rm b} + LF_{\rm L}},  &\; \text{if } b > 0, \\[0.2cm]
\displaystyle
\frac{(1-f)(C_{\rm b}\beta_{\rm b}F_{\rm b} + \alpha_{\rm L}F_{\rm L}^3 + C_{\rm L}\beta_{\rm L}F_{\rm L})}{F_{\rm L}} + \frac{f(L\alpha_{\rm L}F_{\rm L}^3 + C_{\rm L}\beta_{\rm L}F_{\rm L} + C_{\rm b}\beta_{\rm b}F_{\rm b})}{LF_{\rm L}}, & \; \text{otherwise.}
\end{cases}
\label{eq:energyFinal}  
\end{split}
\end{equation}

We can now estimate energy consumption and the performance of any configuration for a given parallel application using our energy model in Equation~\eqref{eq:energyFinal} and performance models in Equations~\eqref{eq:performance}~and~\eqref{eq:performance2}, respectively. Considering fitted models, we can estimate all possible performance and energy consumptions of an application. Then, these estimations are used in the Pareto method to obtain the most optimal performance-energy trade-offs. In the next section, we are going to show the values obtained after the fitting process and the Pareto approach validation on a two-cluster heterogeneous board.

\section{Experimental Evaluation and Results}
\label{sec:Experimental}

This section describes the experimental setup, including the specifications of the platform and the measurement methodology applied to collect data. Then, we will show the hardware design parameters and the values of the parallel fractions obtained through the non-linear regression. We validate our proposed methodology concerning the estimated Pareto frontier against direct measurements. We computed the error of the estimations by using the Mean Absolute Percentage Error (MAPE), defined as:
\begin{equation}
    \text{MAPE} = \frac{1}{n}\sum_{t=1}^{n}\left |\frac{A_t - E_t}{A_t}\right|,
\end{equation}
where $n$ is the number of configurations, $A_t$ is the actual measurement (execution time or energy consumption) and $E_t$ is the estimated execution time or energy consumption.

\subsection{Experimental Setup}

The proposed methodology was validated using an ODROID-XU3~\cite{Hardkernel} board developed by Hardkernel co. with two core types of the same single-ISA, so both can execute the same compiled code. It uses a Samsung Exynos5422 System on a Chip (SoC), which utilizes ARM big.LITTLE technology and constitutes an ARM Cortex-A15 (big) quad-core cluster and a Cortex-A7 (LITTLE) quad-core cluster. The ODROID-XU3 has 19 available frequency levels ranging from 200~MHz up to 2 GHz on Cortex-A15 and 14 available frequency levels from 200 MHz to 1.5 GHz on Cortex-A7. Therefore, considering that each configuration is a combination of the number of cores in each cluster and a frequency pair that is available as a resource for an application to be executed, there are 6384 possible configurations to explore. To further clarify, consider that each cluster can select none to four~cores, then~$(\text{\# of big cores})\times(5\times19) \times (\text{\# of LITTLE cores})\times(5\times14) = 6650 $. For each combination of active numbers of LITTLE and big cores, there are $19\times14$ arrangements of frequencies to choose. Since the combination of $0$ big cores and $0$ LITTLE cores is not possible, the number of available configurations is $6650 - (19\times14) = 6384$. The experimental platform was set up with a Ubuntu Minimal 18.04 LTS running Linux Kernel LTS 4.14.43, and it did not permit disabling individual cores. Our software project can be found in the GitLab {(\url{https://gitlab.com/lappsufrn/XU3EM})}.

The ODROID-XU3 has four current and voltage sensors to measure the power dissipation of the cortex-A15 cores, cortex-A7 cores, GPUs and DRAMs individually. We sampled the clusters' power consumption readings every 0.5 seconds with timestamps. Then, by integrating the power~samples, we computed the energy used by a given application. We consider the average power calculated by dividing total energy by execution time. Moreover, we used \textit{\texttt{cpuset}}~\cite{cpuset} subsystem  to confine all the data collection instrumentation and operating system's processes/threads to the not active cores, when available, and assign the workload application appropriately to the cores. Thus, this mitigates the interference of all non-target application in the experiment.

\subsection{Hardware Design Parameters} 
\label{sub:hardware}
This section shows the values obtained for the parameters that represent the hardware design characteristics. As these parameters are related to the CPU design and they must be application-agnostic, we used a CPU stress tool, called  \texttt{stress-ng}~\cite{stressng}. This tool can load and stress a computer system in various selectable ways and is available in any Linux distribution. This tool has an extensive range of specific stress tests that exercise floating-point, integer, bit manipulation, and control flow operations on the CPU.

\begin{figure}[H]
    \centering
    \includegraphics[scale=0.33]{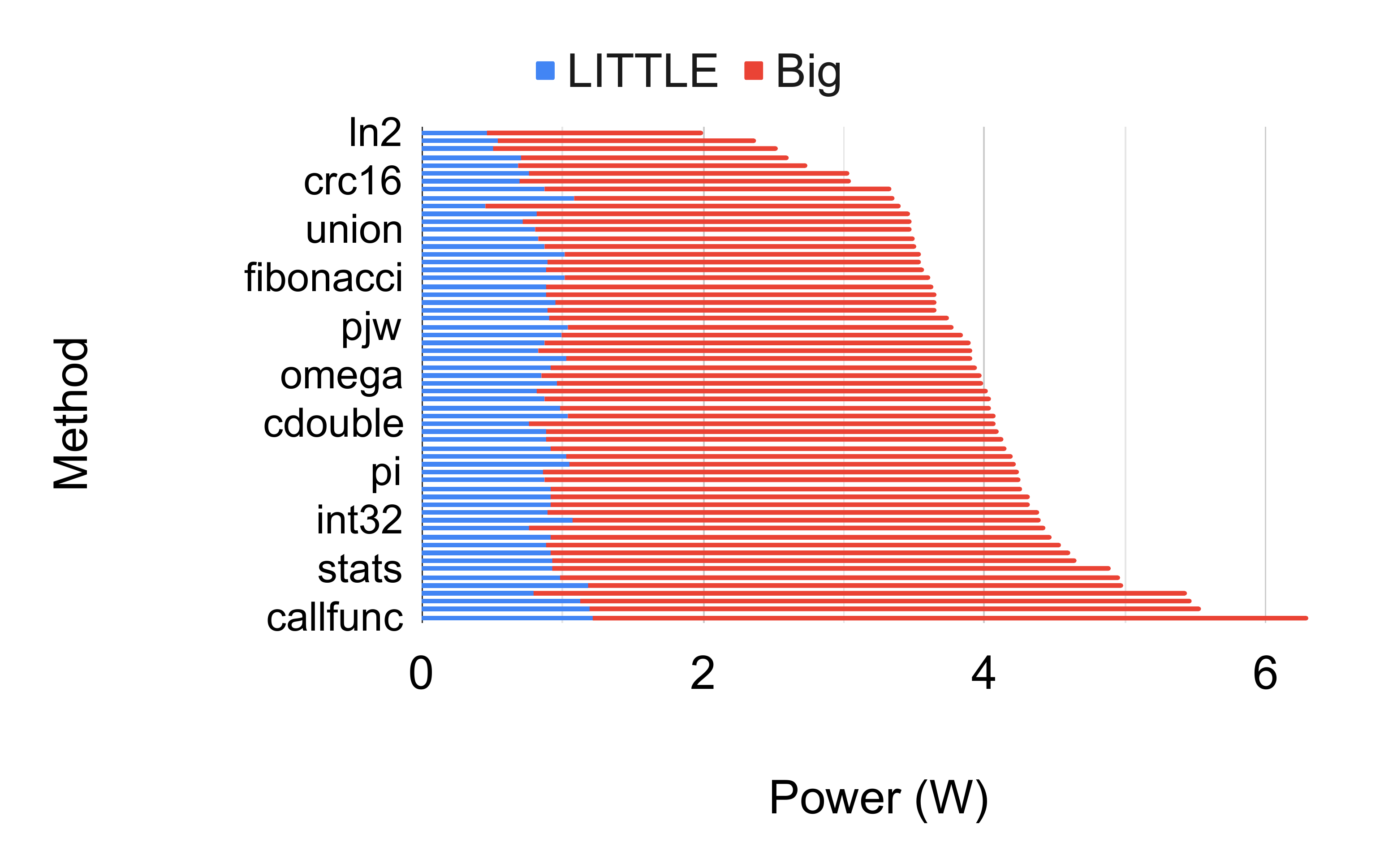}
    \caption{Power consumption from LITTLE and big clusters for every \texttt{stress-ng} method.}
    \label{fig:methods}
\end{figure}

We tested each method to find the one that maximizes the stress on the ODROID-XU3, i.e., the method that led to the highest power draw as cores executed their instructions (see Figure~\ref{fig:methods}). Hence, after executing every method for 15 s on two big cores (1.9 GHz), and four LITTLE cores (1.5 GHZ), we chose the \texttt{callfunc} stress method. Using all four big cores at the highest frequency would have overheated the board. High CPU frequencies, especially from the big cores, produce a large amount of~heat, making the system protection mechanism against high temperatures halt the device to prevent~damage. Moreover, running 15 seconds provides 30 power reading samples from the sensors. The average standard deviations regarding all methods are 0.0792 and 0.2395 for LITTLE and big power consumption, respectively.

\subsubsection{The Perf Parameter} 
\label{sub:perf}
The \textit{\texttt{\emph{perf}}}  performance parameter represents the big core speedup in comparison with the LITTLE core. It is application-agnostic, relying only on the chip design. We executed the \texttt{stress-ng} with \texttt{callfunc} method on one core of each cluster type, fixing the number of operations at 87,555. This~number of operations provides approximately one second to conclude in one big core with the highest frequency and up to 15 s for one LITTLE core with the lowest frequency.

\begin{figure}[H]
    \centering
    \includegraphics[scale=0.45]{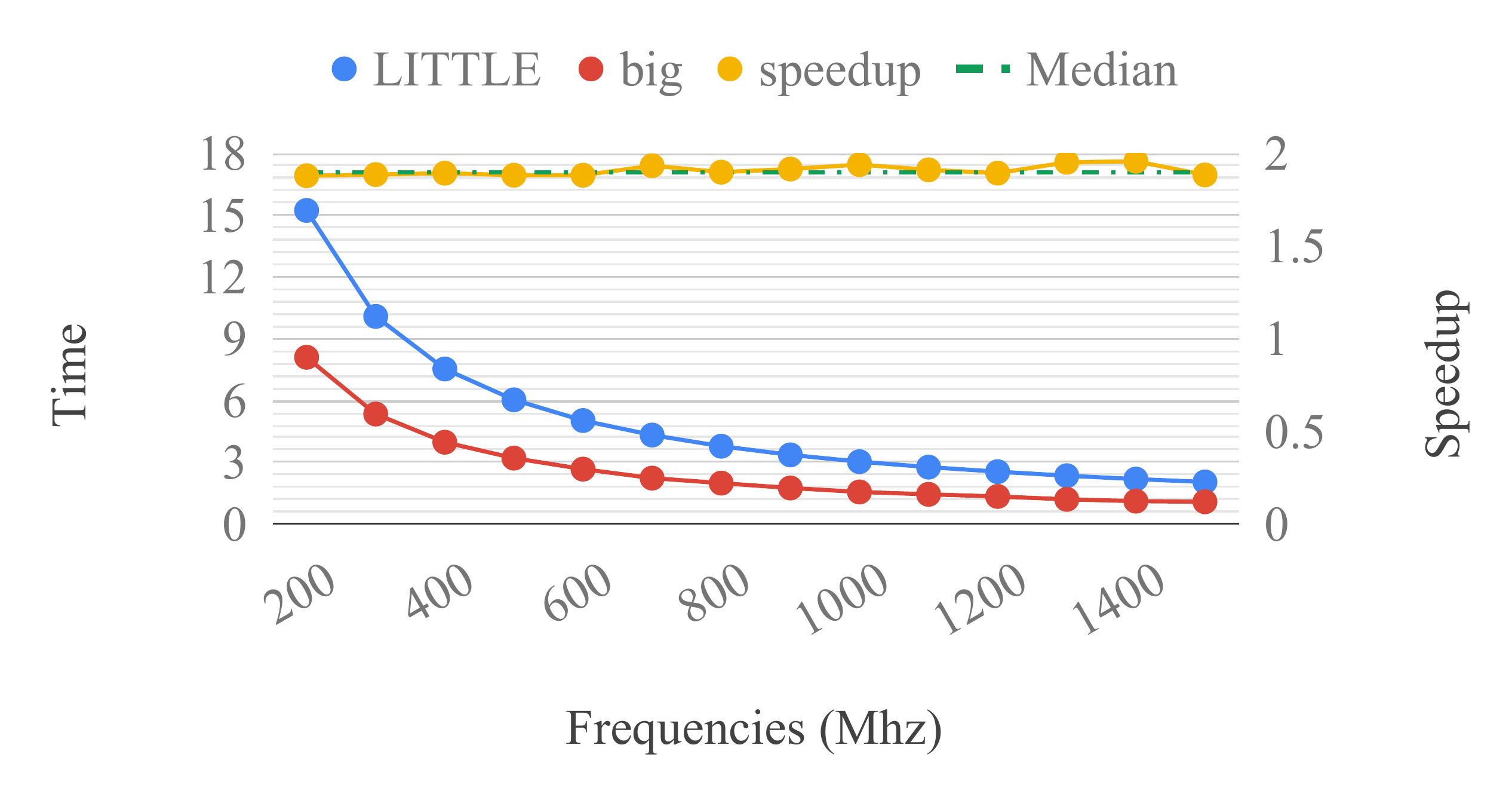}
    \caption{Performance parameter \textit{\texttt{\emph{perf}}} fitted for the architecture using a fixed number of operations of the \texttt{stress-ng} tool for one core of each cluster.}
    \label{fig:perf}
\end{figure}

Figure~\ref{fig:perf} shows the result of \textit{\texttt{\emph{perf}}} fitting. The left-hand side Y-axis shows the median execution time of five runs for one core of each cluster type and different frequencies. The frequency range used is 200 MHz--1.5 GHz. Higher values are not possible due to the maximum LITTLE core frequency. The~right-hand side Y-axis represents the speedup of one big core when comparing to one LITTLE core for each frequency. The average standard deviations of the runs regarding all frequencies are 0.0252 and 0.0404 for LITTLE and big core, respectively. The median of every speedup is $1.897$ which is the obtained \textit{\texttt{\emph{perf}}} value.

\subsubsection{Power Consumption Parameters}
 \label{sub:power}
We aim to obtain a power model that independent of the application, represents the power consumption of the two-clusters. As we want to model the total power consumption of the entire chip, we measured the whole chip to obtain $P_{\rm HMP}$. We defined the number of processes and which cores would be active to execute the \texttt{stress-ng} by generating 95 evenly distributed Halton configurations. Each configuration is composed of the number of active big cores $b$, active LITTLE cores $L$, big's frequency cluster $F_{\rm b}$ and LITTLE's frequency cluster $F_{\rm L}$. 

We used the \texttt{stress-ng} with \texttt{callfunc} stress method and a fixed execution time of 75 seconds, with the command line {\tt stress-ng --cpu $N$ --cpu-method callfunc -t 75 --taskset $t$.} The \texttt{--cpu} parameter determines the number of processes on which the same method will be executed, 
and the parameter \texttt{--taskset} can be used to set a process's CPU affinity.

Running 75 seconds provides 150 power samples. The average standard deviations regarding all 95 configurations are 0.0164 and 0.1478 for LITTLE and big core power consumption, respectively. As the configuration parameters and their total power consumption measurements are known, we~fit the parameters $\alpha_{\rm L}, \beta_{\rm L},\alpha_{\rm b},\beta_{\rm b}$ from the Equation~\eqref{eq:powerHMP}. {The fitted values are}: $\alpha_{\rm L} = 5.953 \times 10^{-29}, \alpha_{\rm b} = 2.914 \times 10^{-28}, \beta_{\rm L} = 1.033 \times 10^{-10},$ $\beta_{\rm b} = 9.342 \times 10^{-11}$. 

\subsection{Applications}
\label{sub:app}
To validate our approach, we used eight applications: the \texttt{Black-Scholes}, \texttt{Bodytrack}, \texttt{Freqmine} applications provided by PARSEC,  \texttt{Smallpt} and \texttt{x264} by Phoronix Test Suite and~\texttt{kmeans}, \texttt{Particle Filter} and \texttt{LavaMD} from Rodinia benchmark. PARSEC~\cite{Bienia2011} is a well-known benchmark suite of parallel applications, which focuses on emerging workloads and is designed to be representative of next-generation shared-memory programs for chip-multiprocessors. The Phoronix Test Suite~\cite{Phoronix} is a benchmarking platform that provides an extensible framework for carrying out tests in a fully automated manner from test installation to execution and reporting. Rodinia benchmark~\cite{Che2009} is a set of applications designed for heterogeneous computing infrastructures with OpenMP~\cite{OpenMP}, OpenCL~\cite{OpenCVteam}, and CUDA~\cite{Corporation2016} implementations. 

We have chosen the applications because they were implemented using OpenMP. This allowed us to modify their workloads to be dynamically balanced using dynamic OpenMP schedule clause~\cite{openmpclause}. Moreover, the number of threads created is equal to the number of active cores available of a given application under a configuration. The thread master was bound to one big core in order to make sure that the sequential part of the code would run on the higher performance~core. The~other threads were bound to one core each. This mean that these applications follow the assumptions we made for the performance model. 

The notable exception for the chosen criteria is the \texttt{x264} encoder provided by the Phoronix test suite since it has only POSIX threads for parallelism. As higher video quality significantly depends on the encoding process, i.e., longer execution time and higher energy consumption, it is essential to improve the energy efficiency of video encoding on embedded devices. We therefore evaluate the behaviour of our methodology on this video encoder.

The characteristics of the applications which are used in this paper, according to~\cite{Bienia2011,Che2009,Goodrum2012,Ozisikyilmaz2006,Che2010}, can be seen in Table~\ref{tab:appDesc}. Most applications are CPU and memory intensive, so we expect a considerable use of the HMP chip and the memory system as well as the communication between them.

Nevertheless,  some implicit hardware or software features, such as communication-to-computation ratio, cache sharing and off-chip traffic, are not explicitly included in our power and performance model. What we want to demonstrate is that a low-overhead, straightforward model is sufficient to estimate a Pareto frontier. Besides, as those features should limit the parallel speedup, it is expected that the sequential fraction of the application may be more significant for some applications.

We strive to capture the nuances of each configuration that will affect the way the application works on the hardware, producing different execution time. Therefore, we fit the performance model to find the specific parallel fraction $f$ of each application by executing 50 Halton configurations ($F_{\rm b},F_{\rm L},b,L$).   As Halton generates configurations that cover the domain more evenly than pseudo-random algorithms,  50 configurations  provide sufficient data for us to build our performance model. One more configuration is needed to find $T_{\rm L}(F)$, it is $b=$~0, $L=$~1, $F_{\rm b}=$~2 GHz, $F=F_{\rm L}=$~800~MHz. We calculated the median of the execution time of five runs for each configuration.

Considering the fitted \textit{\texttt{\emph{perf}}}, and the Equations~\eqref{eq:performance} and~\eqref{eq:performance2}, we use non-linear regression to find the parallel fraction $f$ for every application, as explained in the Section~\ref{sec:methodology}. Table~\ref{tab:fittingerrors} shows the value of $f$ for each application. Note that \texttt{Black-Scholes} and \texttt{kmeans} have a parallel fraction of 0.7743 and 0.6381, respectively, indicating a larger sequential portion of these applications when compared to other applications. The following section will show the validation of the Pareto frontier. It provides the evidence that our low-overhead models are sufficient for achieving our goals.

\newpage
\paperwidth=\pdfpageheight
\paperheight=\pdfpagewidth
\pdfpageheight=\paperheight
\pdfpagewidth=\paperwidth
\newgeometry{layoutwidth=297mm,layoutheight=210 mm, left=2.7cm,right=2.7cm,top=1.8cm,bottom=1.5cm, includehead,includefoot}
\fancyheadoffset[LO,RE]{0cm}
\fancyheadoffset[RO,LE]{0cm}

\begin{table}[H]
\centering
\tablesize{\footnotesize} 

\caption{C{haracteristics detail of each application based} on~\cite{Bienia2011, Che2009, Goodrum2012, Ozisikyilmaz2006, Che2010}.}
\label{tab:appDesc}
\scalebox{.7}[0.7]{\begin{tabular}{llllll}
\toprule
\textbf{Benchmark} & \multicolumn{1}{c}{\textbf{App.}} & \multicolumn{1}{c}{\textbf{Domain}} & \multicolumn{1}{c}{\textbf{Type}} & \multicolumn{1}{c}{\textbf{Problem Size}} & \multicolumn{1}{c}{\textbf{Description}} \\ \midrule
\multirow{3}{*}{\textbf{PARSEC}} & \texttt{Black-Scholes} & Financial Analysis & CPU intensive & 10,000,000 options & Portfolio price calculation using Black-Scholes PDE \\ \cmidrule{2-6} 
 & \texttt{Bodytrack} & Computer Vision & CPU and memory intensive & 4 cameras, 261 frames, 4000 particles, 5 annealing layers & Computer vision, tracks 3D pose of human body \\ \cmidrule{2-6} 
 & \texttt{Freqmine} & Data \mbox{Mining} & CPU and memory intensive & Database composed of spidered collection of 250,000 web HTML documents & Array-based version of the FP-growth method \\ \midrule
\multirow{3}{*}{\textbf{Rodinia}} & \texttt{Kmeans} & Data \mbox{Mining} & CPU and memory intensive & 1000,000 points, 34 dimensions,  5 clusters & Mean-based data partitioning method \\ \cmidrule{2-6} 
 & \texttt{Particle Filter} & Medical Imaging & CPU and memory intensive & Video resolution 128 $\times$ 128, 10 frames, 10,000 particles & Tracks cells by  statisticaly estimating the path in a Bayesian framework. \\ \cmidrule{2-6} 
 & LavaMD & Molecular Dynamics & CPU and Memory intensive & 100 boxes & Computes the interactions between particles in a three-dimensional cubic space \\ \midrule
\multirow{2}{*}{\textbf{Phoronix}} & \texttt{Smallppt} & Image Renderer & CPU intensive & 1024 $\times$ 768, 128 samples/ pixel & Generates a single image of a modified Cornell box rendered with full global illumination \\ \cmidrule{2-6} 
 & \texttt{x264} & Media \mbox{Processing} & CPU and memory intensive & 1920 $\times$ 1080 pixels (HDTV resolution), 600 frames & H.264 video encoder \\ \bottomrule
\end{tabular}%
}
\end{table}
\unskip
\begin{table}[H]
\centering
\caption{The fitted parallel fraction values of each application.}
\label{tab:fittingerrors}
\scalebox{.9}[0.9]{\begin{tabular}{ccccccccc}
\toprule
\textbf{Application} & {\textbf{Black-scholes}} & {\textbf{Bodytrack}} & {\textbf{Freqmine}} & {\textbf{Smallpt}} & {\textbf{x264}} & {\textbf{Kmeans}} & {\textbf{Particle Filter}} & {\textbf{LavaMD}}\\ \midrule
\textbf{$f$} & 0.7743 & 0.9384 & 0.9343 & 0.9898 & 0.888 & 0.6381 & 0.9251 & 0.9961 \\ \bottomrule
\end{tabular}%
}
\end{table}

\newpage
\restoregeometry
\paperwidth=\pdfpageheight
\paperheight=\pdfpagewidth
\pdfpageheight=\paperheight
\pdfpagewidth=\paperwidth
\headwidth=\textwidth

\subsection{Pareto Frontier Validation}
\label{sub:pareto}

In this subsection, we validate our proposed methodology. 
Using the energy model in Equation~\eqref{eq:energyFinal} and the fitted performance model in  Equations~\eqref{eq:performance} and~\eqref{eq:performance2}, we generated a pair of estimated performance and energy consumption values of all possible configurations for each application in Table~\ref{tab:appDesc}.  Then,  the Pareto frontier method selects the performance and energy pairs that offer the optimal trade-off.

For all applications, the Pareto frontier selection step resulted in configurations that use all cores of the two clusters. The frequencies of the cluster big and LITTLE were the only parameters that have changed. This probably occurred because it was not possible to disable the unused cores during the fitting. 
As a result, the energy savings from configurations, which are not using all available cores, are not enough to justify the performance mitigation. Therefore, the optimal configurations become those that use all cores.

\begin{figure}[H]
	\centering
	\subfloat[Black-Scholes Parsec Application.]{\includegraphics[width=0.3\textwidth,keepaspectratio]{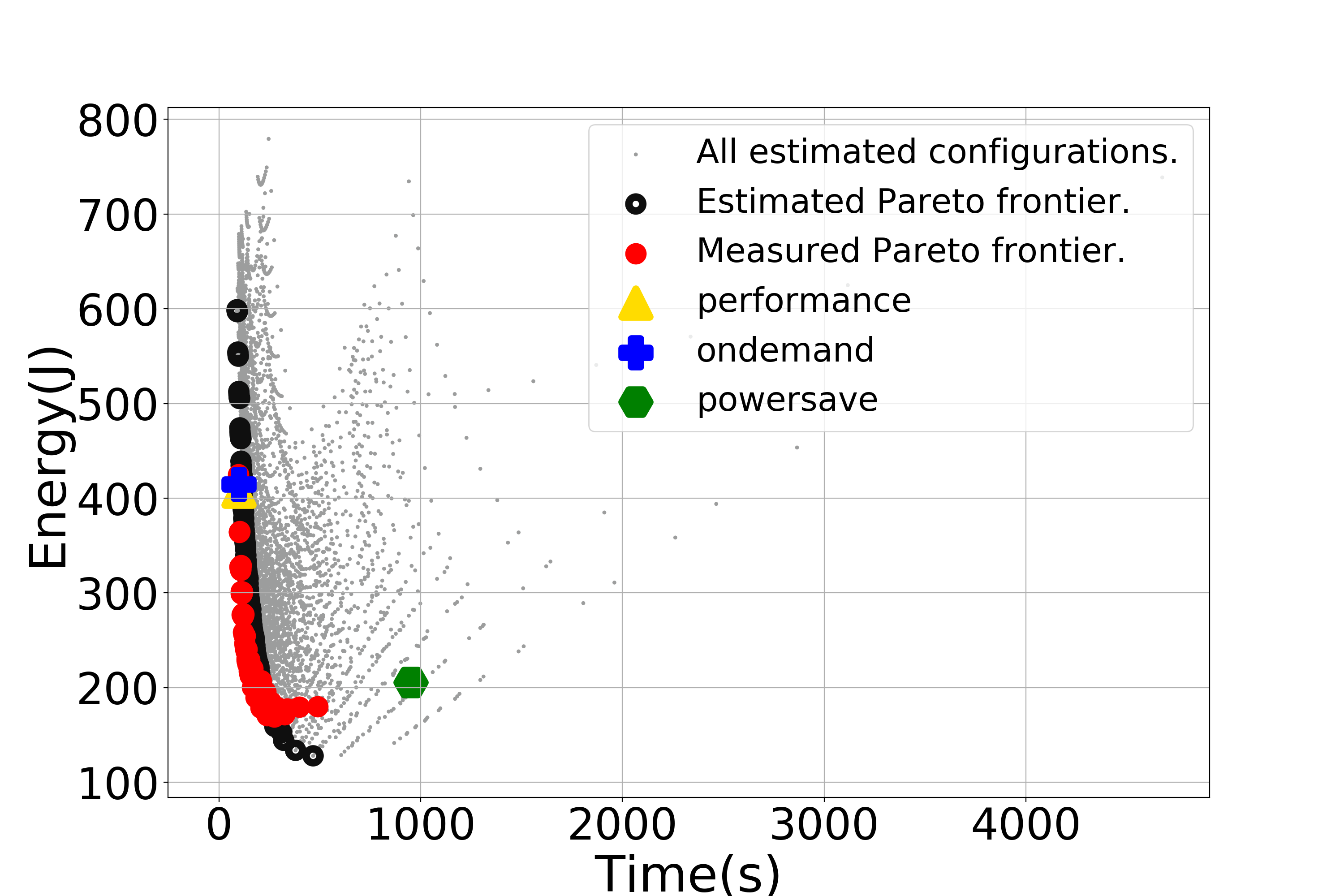}
		\label{fig:black_vali1}}
	\subfloat[bodytrack Parsec Application.]{\includegraphics[width=0.3\textwidth,keepaspectratio]{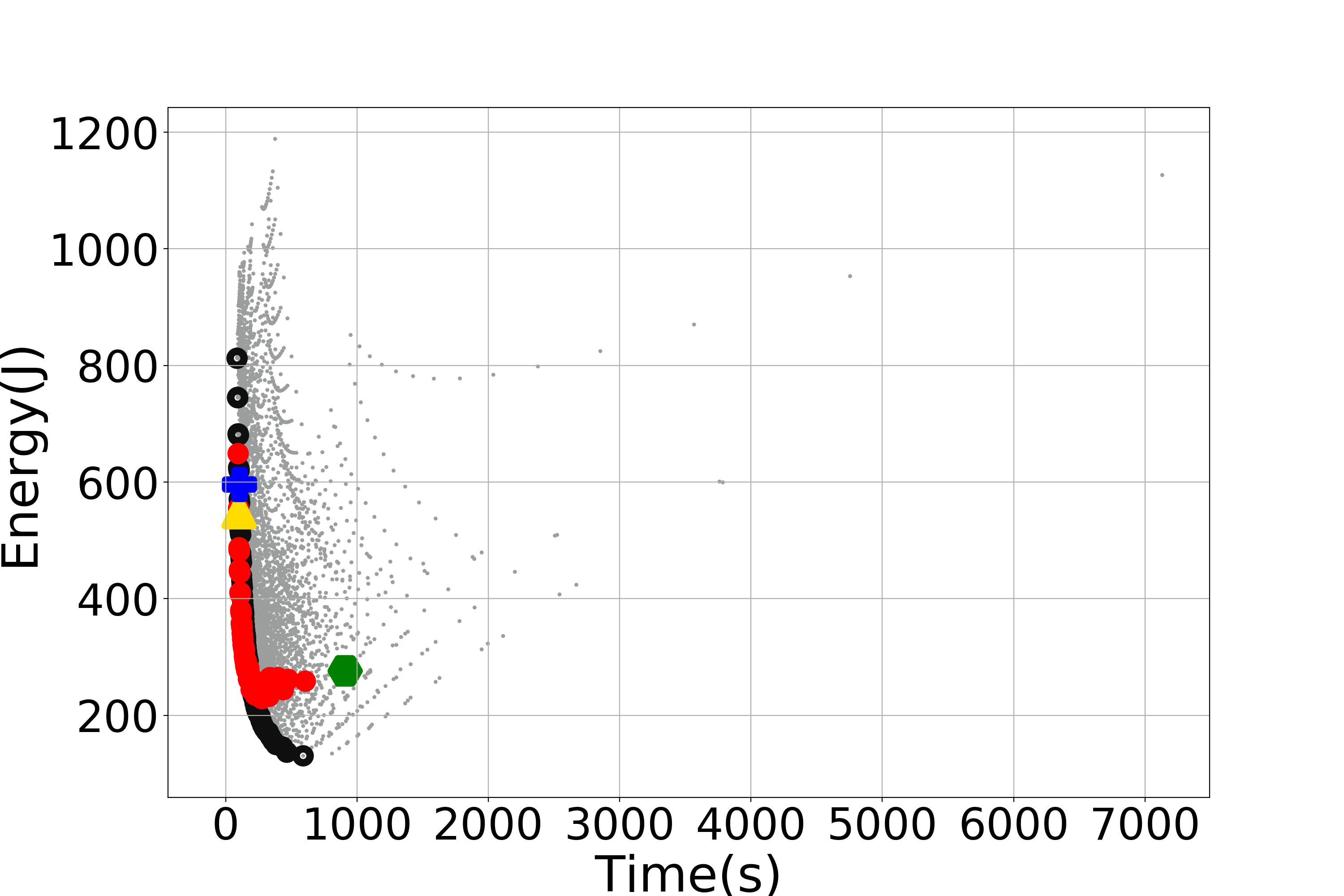}
		\label{fig:body_vali1}}
		
	\subfloat[Freqmine Parsec Application.]{\includegraphics[width=0.3\textwidth,keepaspectratio]{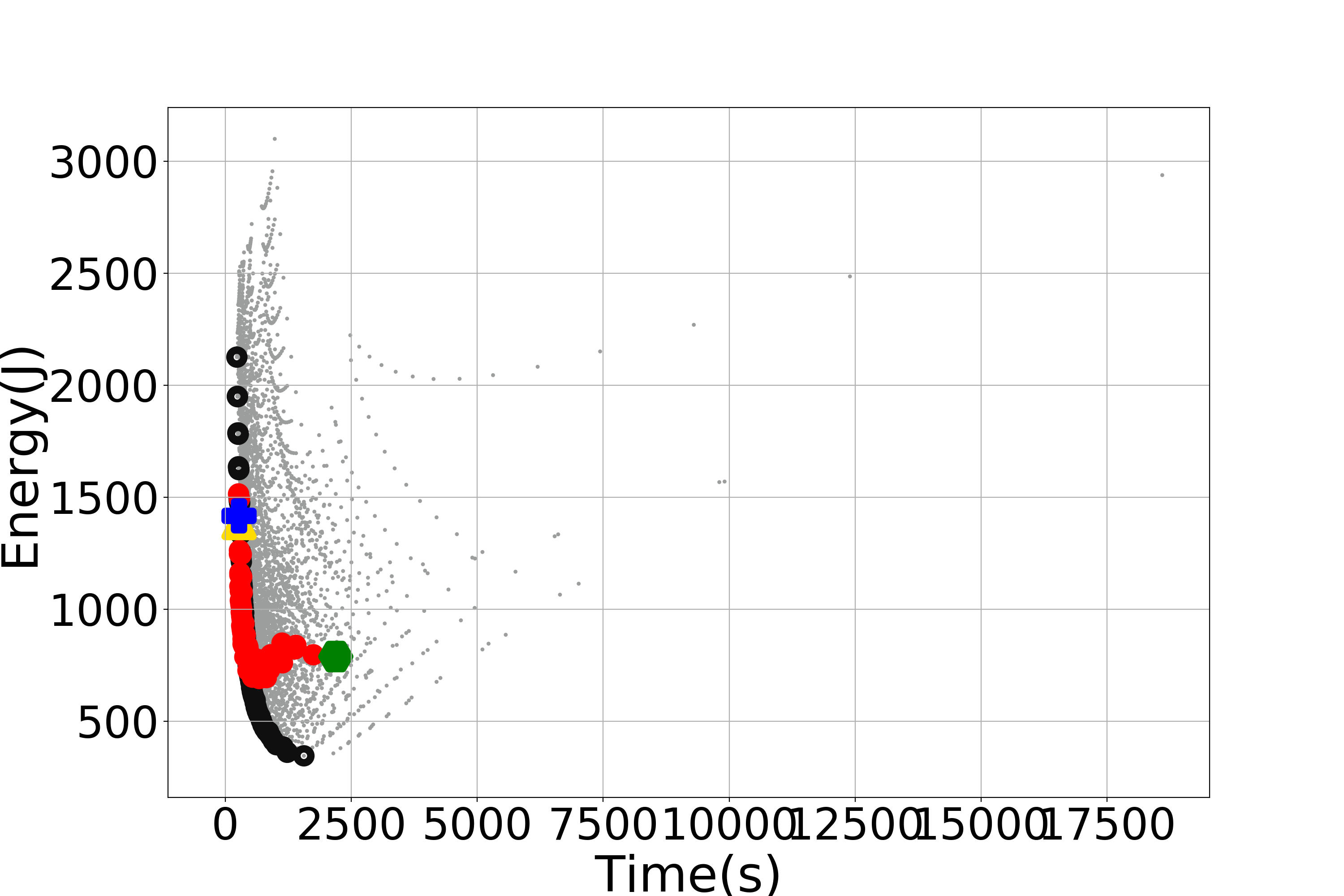}
		\label{fig:freq_vali1}}
	\subfloat[Smallpt Phoronix Application.]{\includegraphics[width=0.3\textwidth,keepaspectratio]{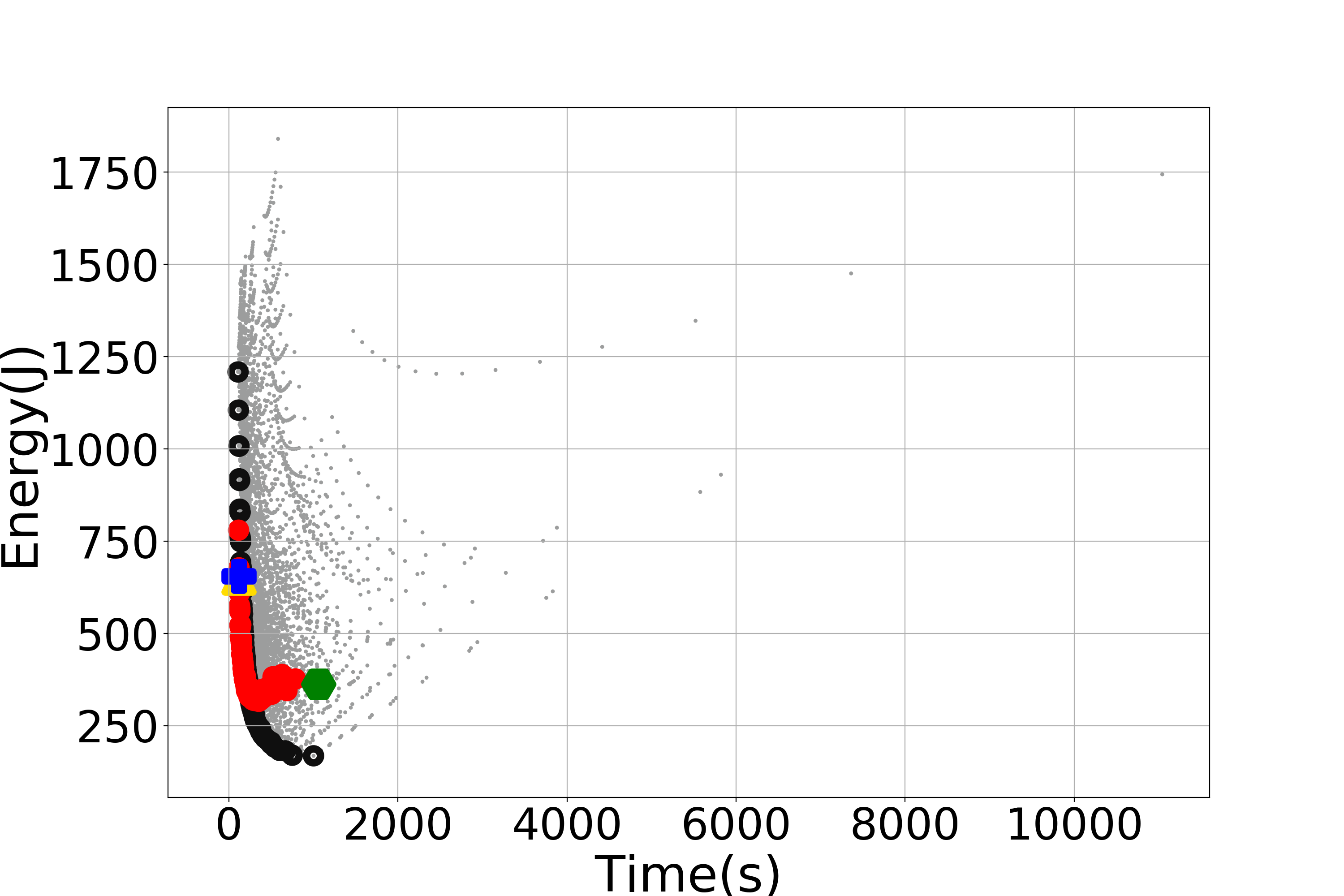}    
		\label{fig:small_vali1}}		

	\subfloat[x264 Phoronix Application.]{\includegraphics[width=0.3\textwidth,keepaspectratio]{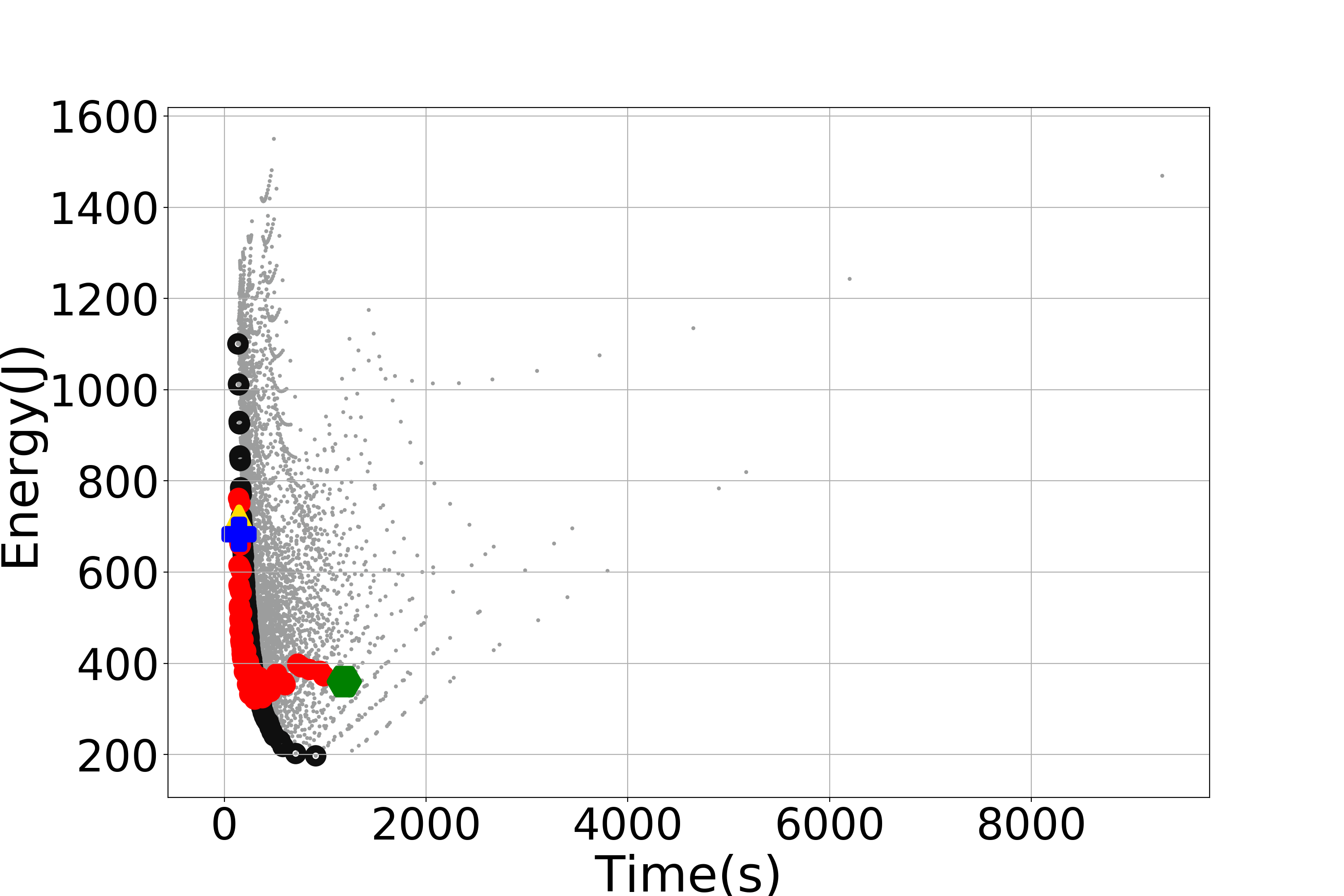}
		\label{fig:x264_vali1}}
	\subfloat[kmeans Rodinia Application.]{\includegraphics[width=0.3\textwidth,keepaspectratio]{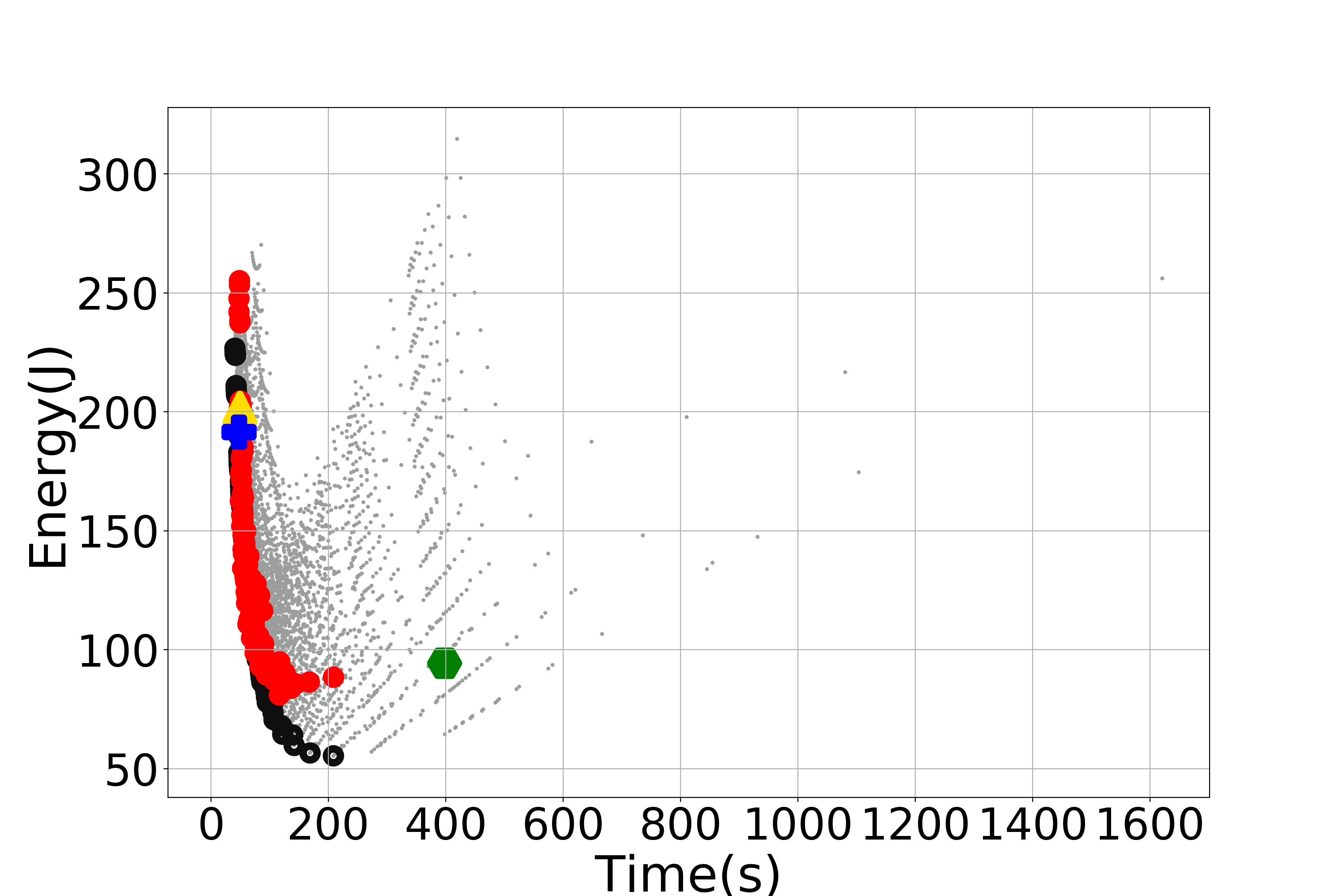}
		\label{fig:kmeans_vali1}}
		
	\subfloat[Particle Filter Rodinia Application.]{\includegraphics[width=0.3\textwidth,keepaspectratio]{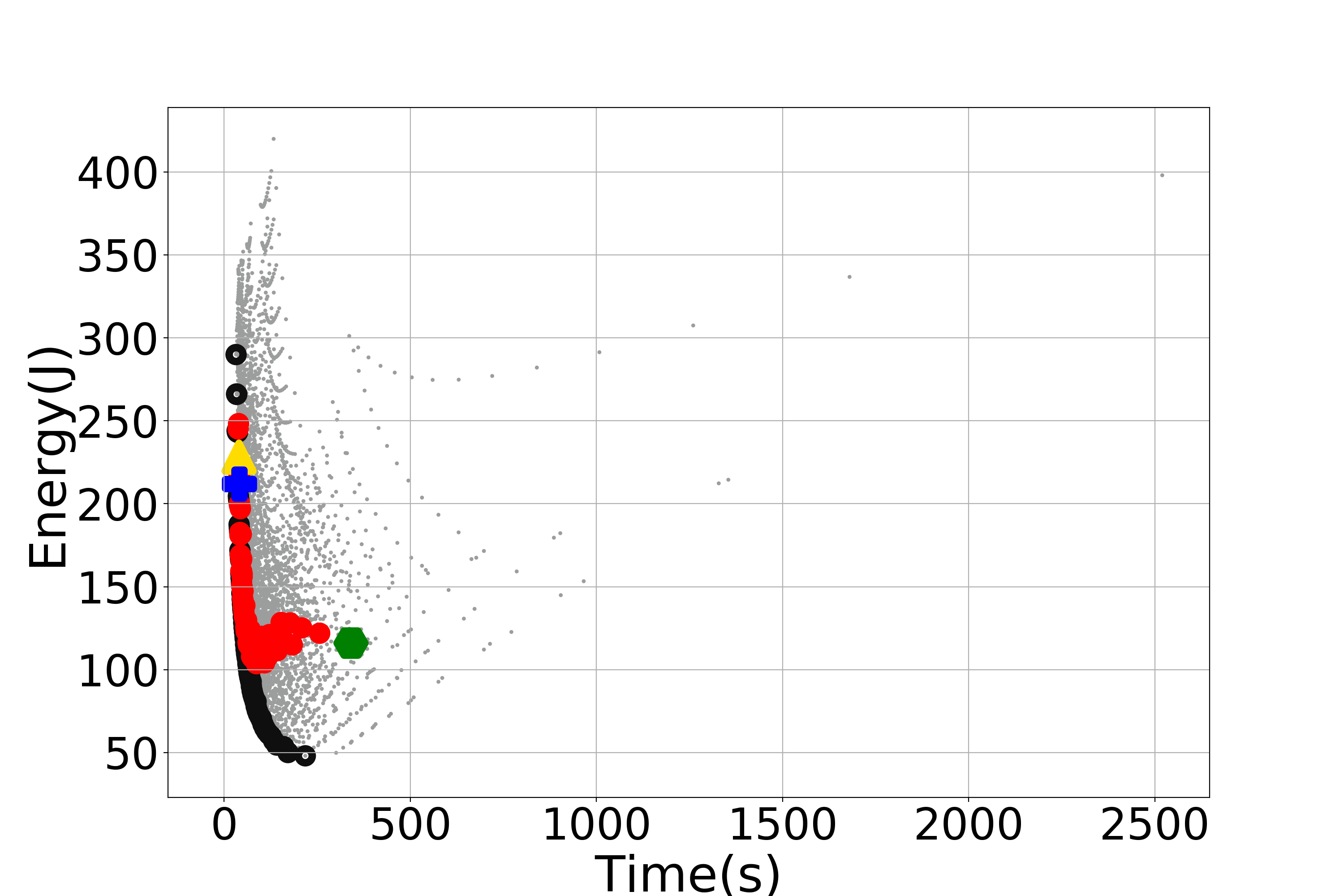}
		\label{fig:part_vali1}}
	\subfloat[LavaMD Rodinia Application.]{\includegraphics[width=0.3\textwidth,keepaspectratio]{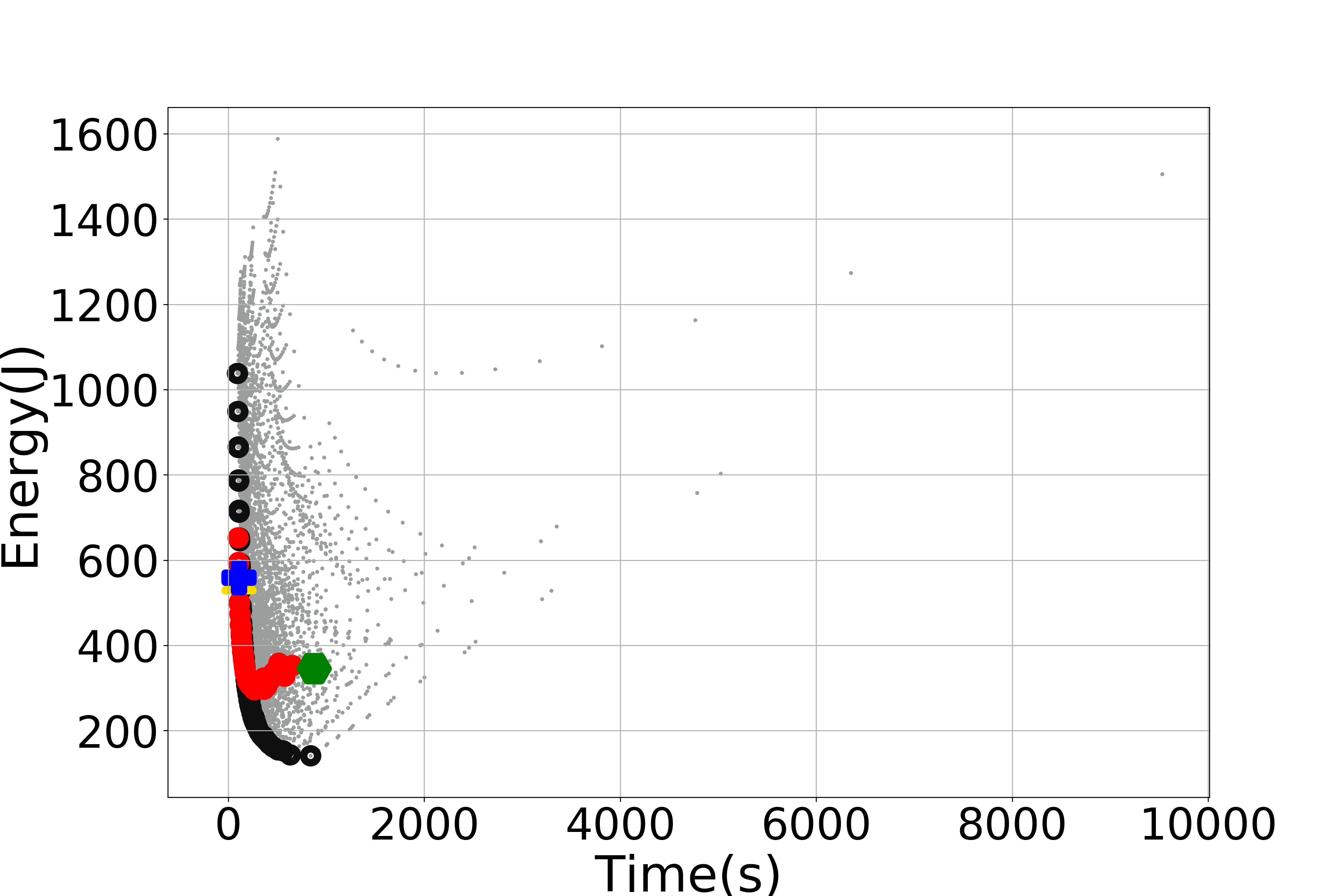}
		\label{fig:lava_vali1}}
	\caption{Estimated and measured Pareto frontier compared with the default governors, and all possible modeled configurations.}
	\label{fig:validation1}	
\end{figure}

Figure~\ref{fig:validation1} shows the estimated and measured Pareto Frontier for all applications w.r.t. all~possible~configurations. Each point describes a pair of performance and energy consumption of a configuration. The grey points are the estimated outcomes of our energy and performance models for all possible configurations. In total, there are 6384 available configurations w.r.t. the combinations of the number of cores of each cluster and pair frequency.

The black circles depict the selected Pareto configurations from the estimated energy and performance pairs. Notice that the estimated Pareto have the least energy consumption concerning different performances. Each red point represents the actual values measured from the sensors using each selected Pareto configuration. Observe that the measured and estimated values are similar to each other but, as the execution time increases, the difference between the measured and estimated energy consumption also rises. The yellow triangle, the blue plus sign and the green hexagon are the \texttt{performance}, \texttt{ondemand} and \texttt{powersave} Linux governors, respectively. We compare our methodology with them in the next section.

Table~\ref{tab:variation} shows the total number of configurations selected by Pareto Frontier and the measured Pareto Frontier variations of energy and performance for each application. Our approach decreased the search space of 6384 available configurations for the given platform by approximately 99\%. 
Considering the highest and least energy consumption and performance among the measured Pareto configurations, we observe energy savings of up to 68.23\% and performance gains of up to 75.62\% for kmeans application. The \texttt{Freqmine} application presented the least variation of energy, up to 54.35\%, however, the performance gains were up to~85.41\%. This demonstrates that the selected configurations by Pareto frontier can provide a good range of performance improvement and energy saving options for operating systems.

\begin{table}[H]
    \centering
      \caption{Pareto Frontier energy and performance variations concerning the measured values and the total number of Pareto configurations.}
    \label{tab:variation}
        \scalebox{.9}[0.9]{
        \begin{tabular}{lcccccccc}
        \toprule
         & \multicolumn{8}{c}{\textbf{Application}} \\ \cmidrule{1-9} 
        \textbf{} & \rotatebox[origin=c]{90}{\textbf{Black-scholes}} & \rotatebox[origin=c]{90}{\textbf{Bodytrack}} & \rotatebox[origin=c]{90}{\textbf{Freqmine}} & \rotatebox[origin=c]{90}{\textbf{Smallpt}} & \rotatebox[origin=c]{90}{\textbf{x264}} & \rotatebox[origin=c]{90}{\textbf{Kmeans}} & \rotatebox[origin=c]{90}{\textbf{Particle Filter}} & \rotatebox[origin=c]{90}{\textbf{LavaMD}} \\ \midrule
        \multicolumn{1}{l}{\textbf{Number of Pareto configurations}} & 93 & 71 & 72 & 66 & 78 & 108 & 74 & 57 \\ \midrule
        \multicolumn{1}{l}{\textbf{Variation of Energy}} & 60.20\% & 64.71\% & 54.35\% & 59.43\% & {57.74\%} & {68.23\%} & {58.28\%} & {54.48\%} \\ \midrule
        \multicolumn{1}{l}{\textbf{Variation of Performance}} & 80.47\% & 84.84\% & 85.41\% & 89.53\% & 85.55\% & 75.62\% & 83.97\% & 88.58\% \\ \bottomrule
        \end{tabular}%
  }
\end{table}

\begin{figure}[H]
	\centering
	\subfloat[Black-Scholes Parsec Application.]{\includegraphics[width=0.38\textwidth,keepaspectratio]{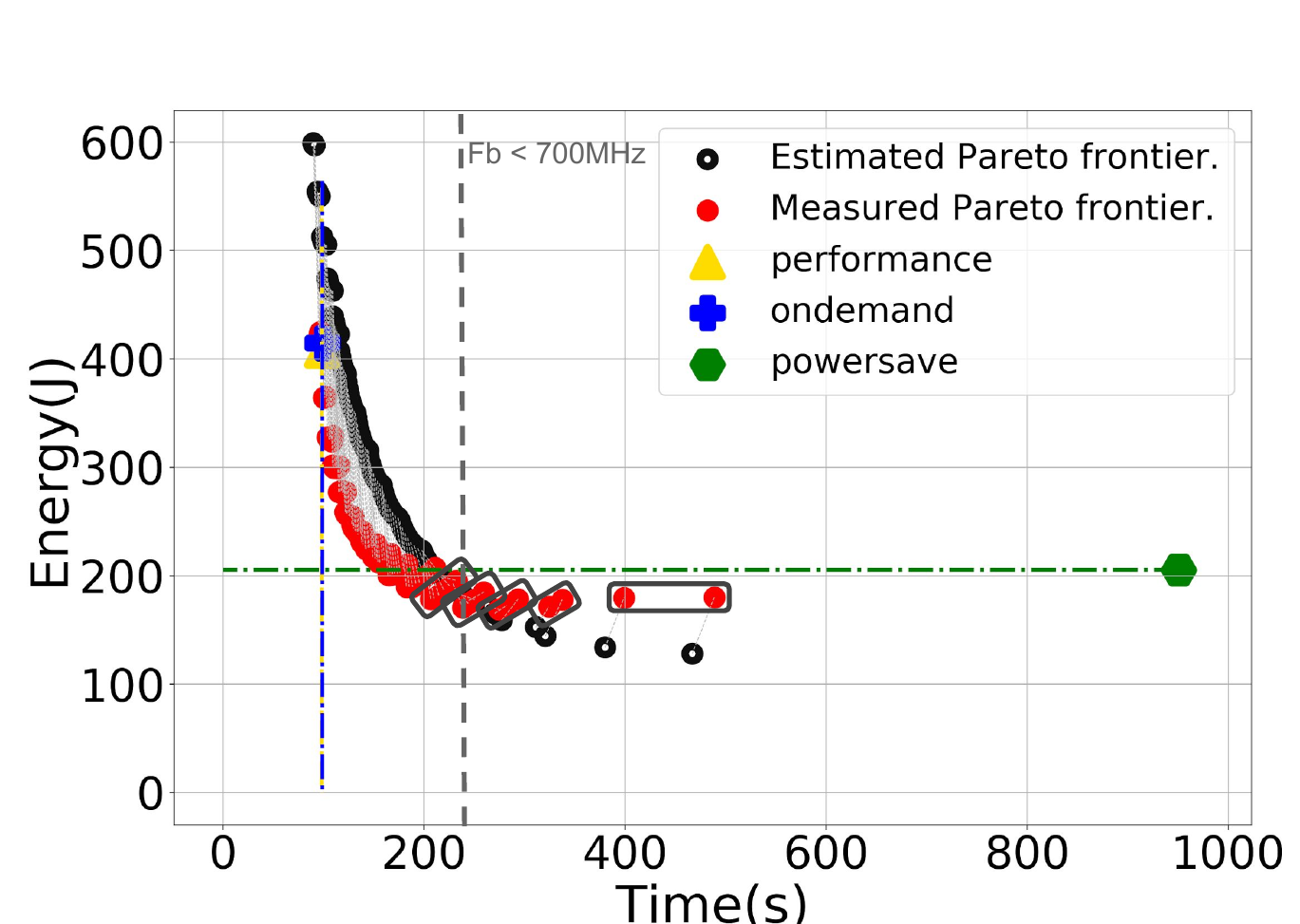}
		\label{fig:black_vali2}}
	\subfloat[bodytrack Parsec Application.]{\includegraphics[width=0.38\textwidth,keepaspectratio]{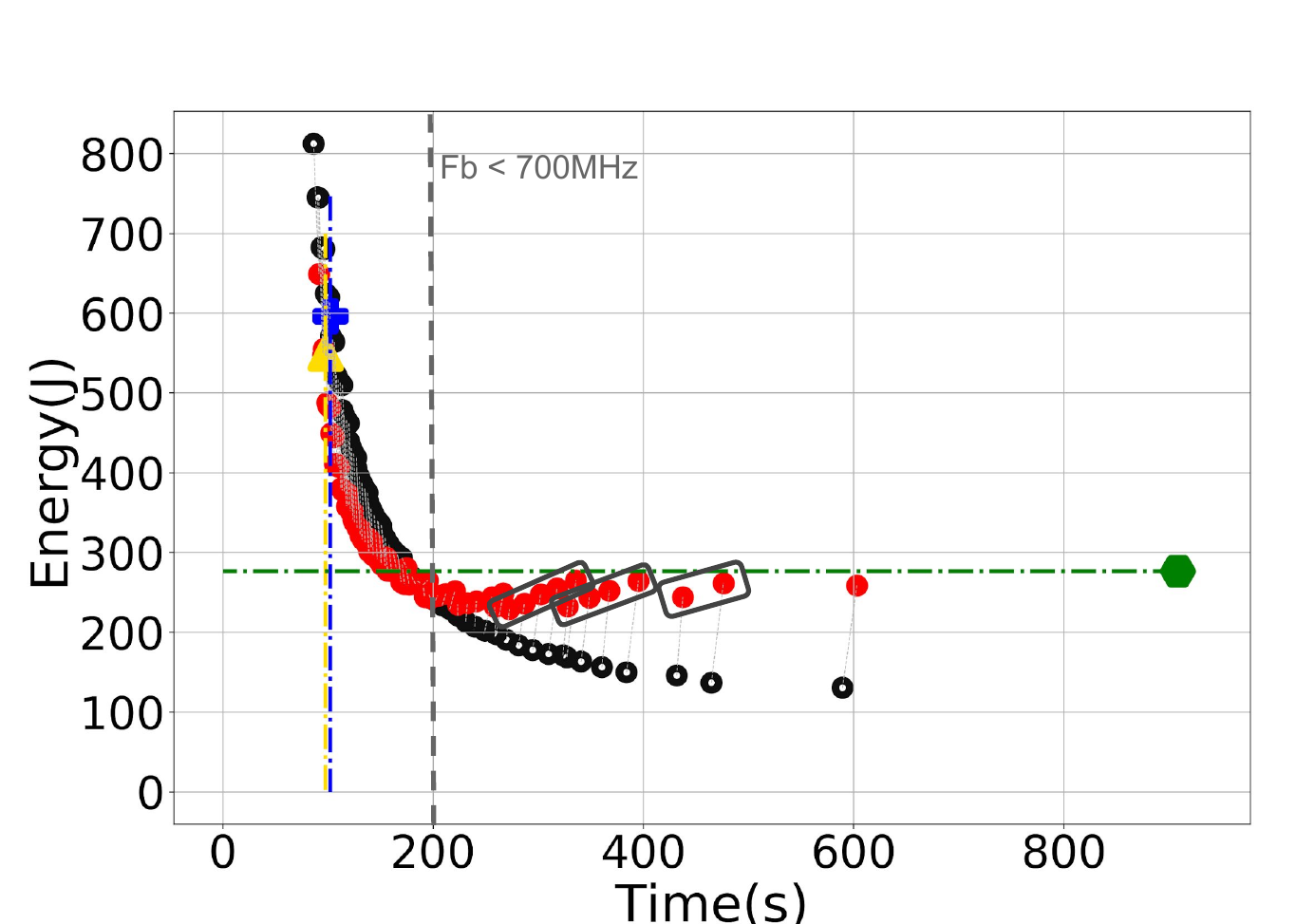}
		\label{fig:body_vali2}}
		
	\subfloat[Freqmine Parsec Application.]{\includegraphics[width=0.38\textwidth,keepaspectratio]{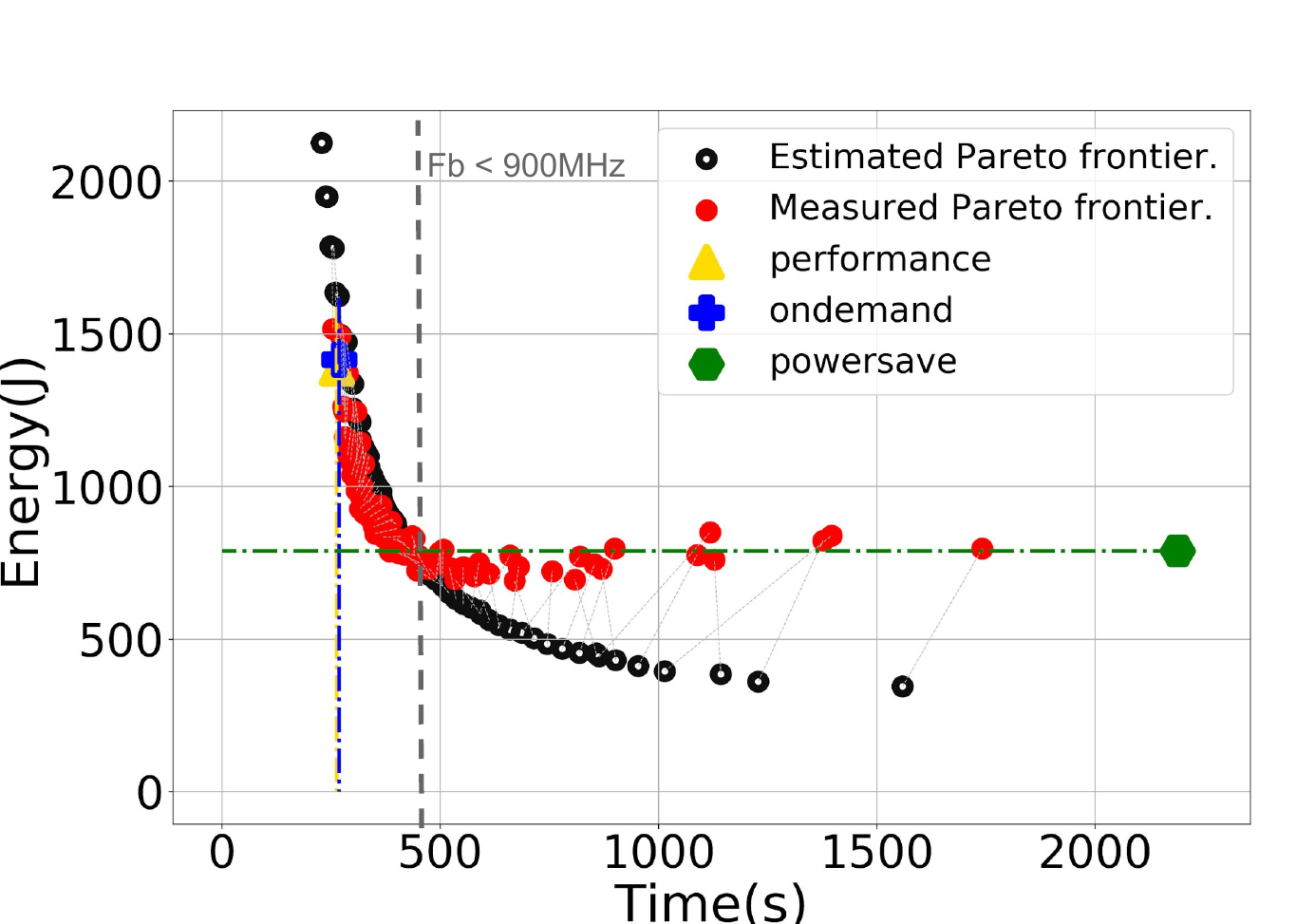}
		\label{fig:freq_vali2}}
	\subfloat[Smallpt Phoronix Application.]{\includegraphics[width=0.38\textwidth,keepaspectratio]{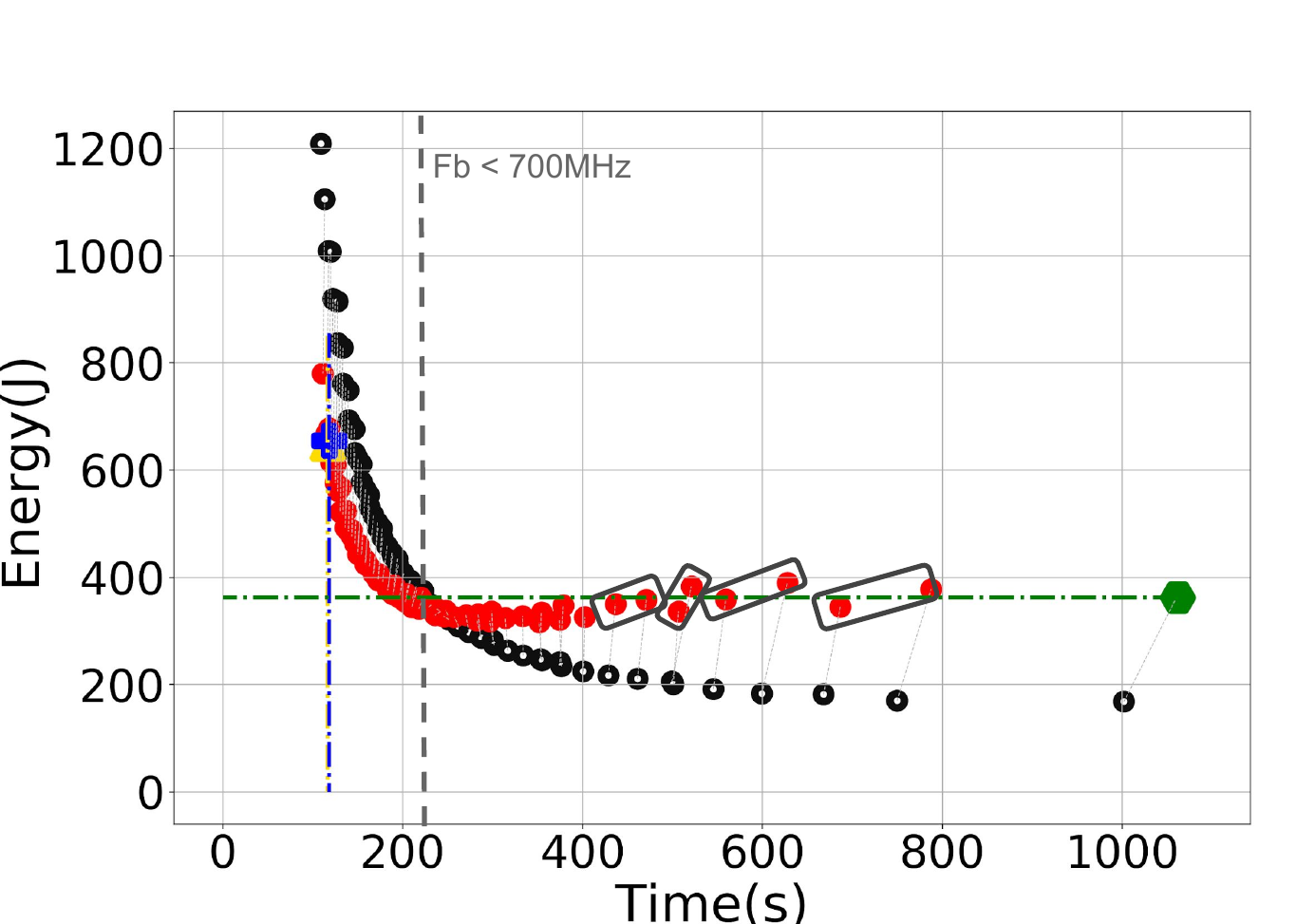}
		\label{fig:small_vali2}}		

	\subfloat[x264 Phoronix Application.]{\includegraphics[width=0.38\textwidth,keepaspectratio]{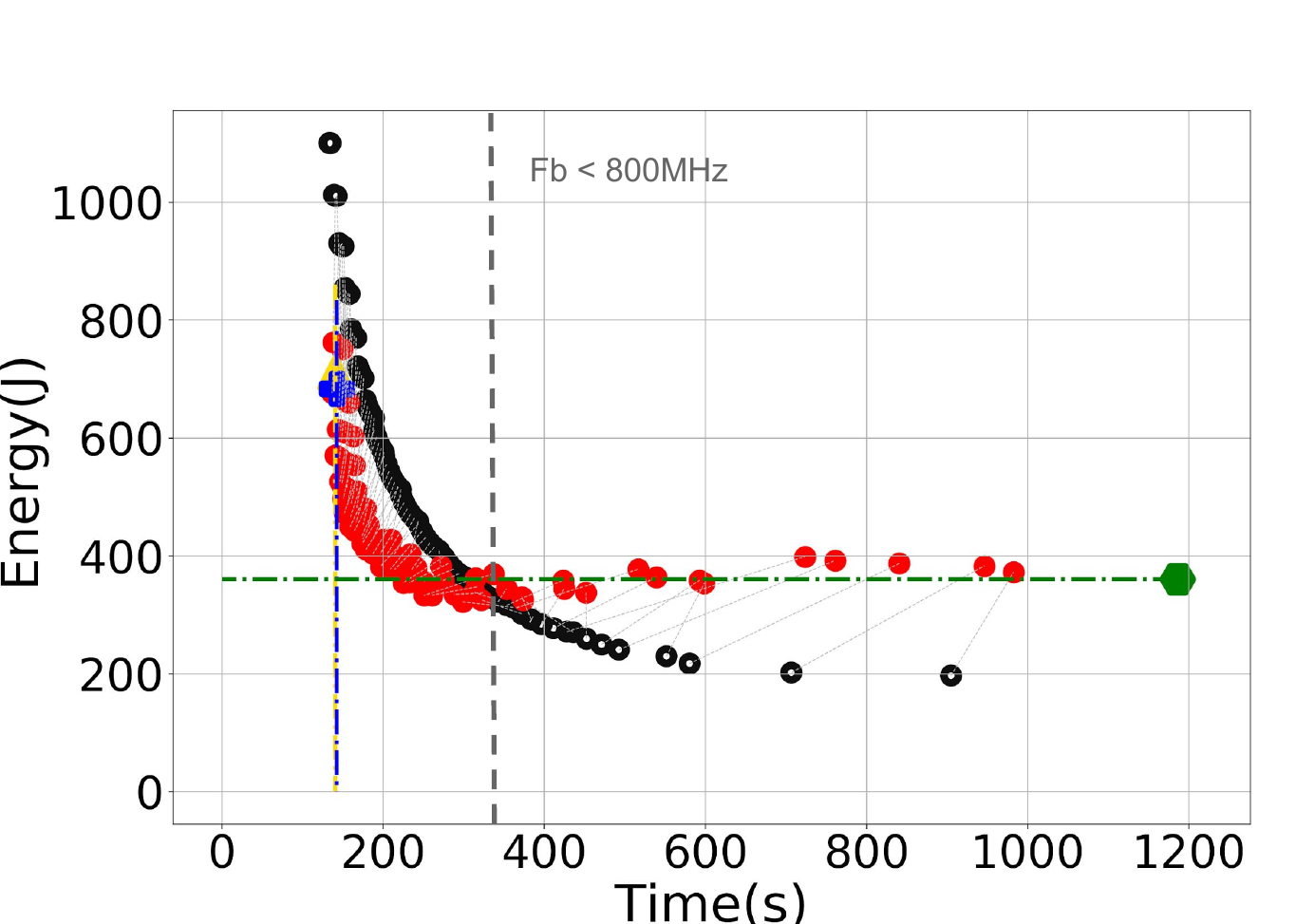}
		\label{fig:x264_vali2}}
	\subfloat[kmeans Rodinia Application.]{\includegraphics[width=0.38\textwidth,keepaspectratio]{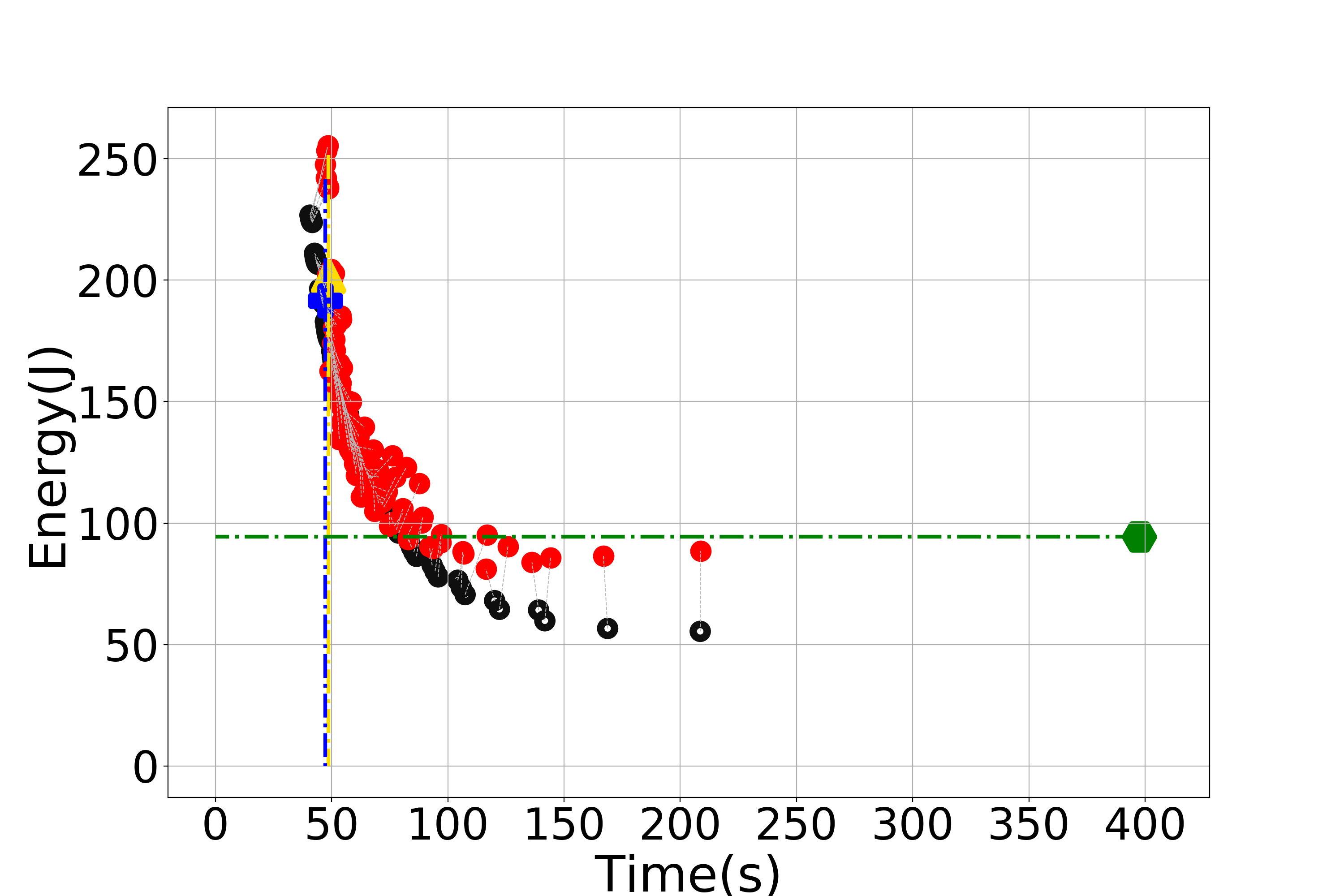}
		\label{fig:kmeans_vali2}}
		
	\subfloat[Particle Filter Rodinia Application.]{\includegraphics[width=0.38\textwidth,keepaspectratio]{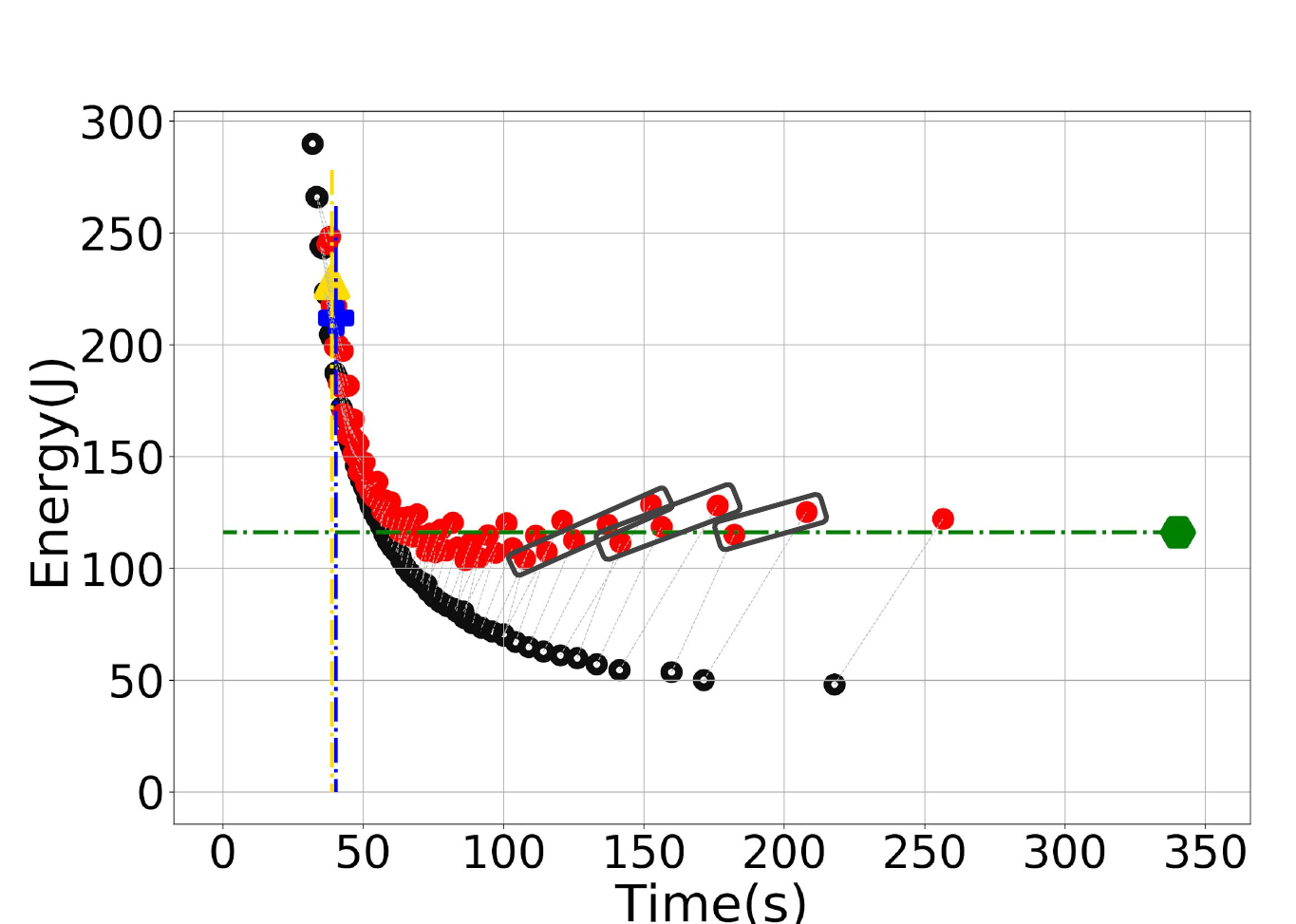}
		\label{fig:particle_vali2}}
	\subfloat[LavaMD Rodinia Application.]{\includegraphics[width=0.38\textwidth,keepaspectratio]{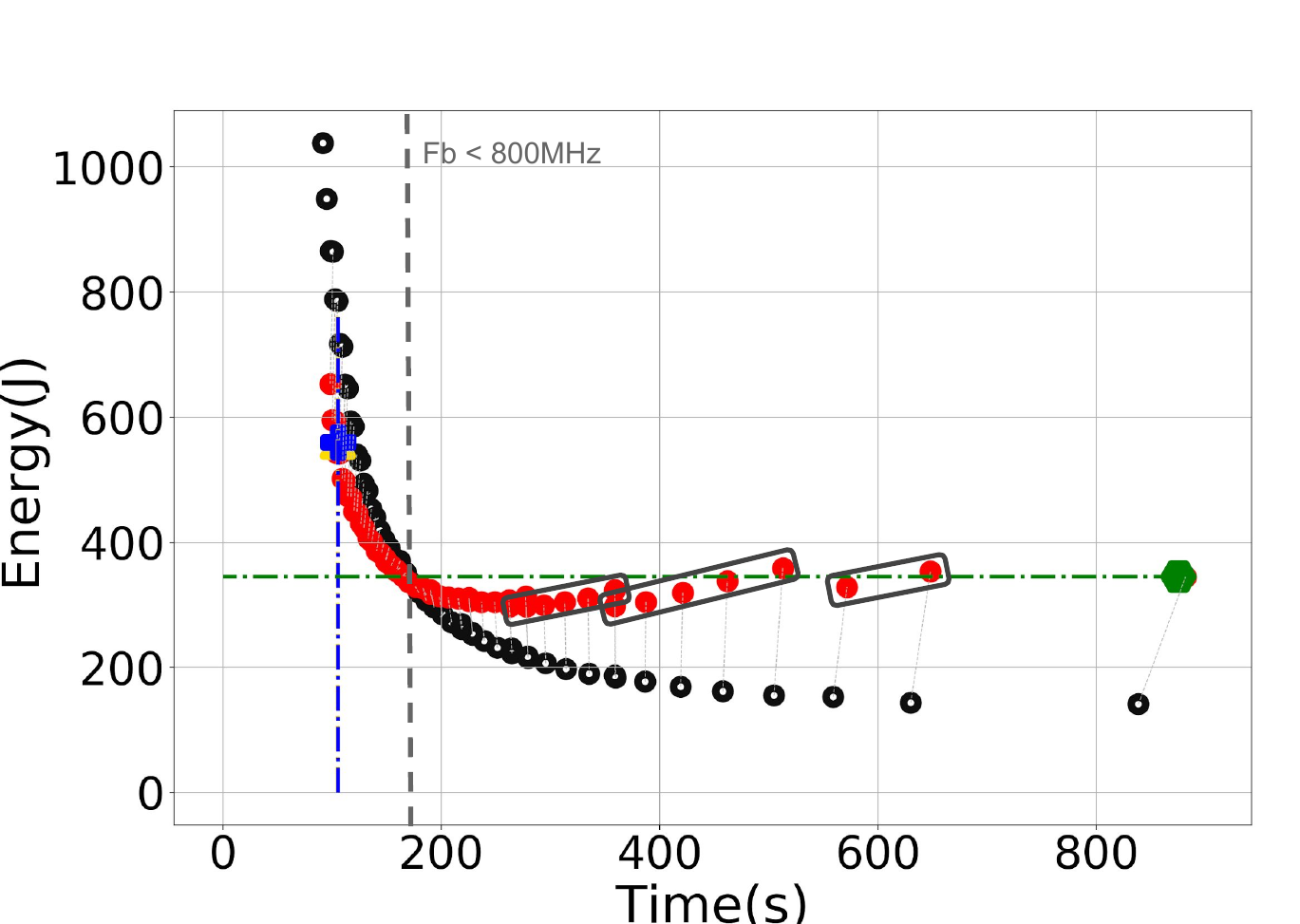}
		\label{fig:lava_vali2}}
	\caption{Estimated and measured Pareto frontier compared with the default governors removing all possible modeled configurations.}
	\label{fig:validation2}	
\end{figure}

Figure~\ref{fig:validation2} gives a zoom-in in Figure~\ref{fig:validation1}. The grey points that represent all possible estimated configurations' outcomes are not displayed. Moreover, the measured and the related estimated Pareto configurations are linked using a dotted line, thus, showing the configurations' results from the models against the measured data.
Depending on the application, we could not execute some configurations to avoid overheating the board. As anticipated in Section~\ref{sub:hardware}, high CPU frequencies produce high temperatures that make the system halt. For instance, the \texttt{Freqmine} results presented in Figure~\ref{fig:validation2}c show four estimated Pareto points with the frequencies (big,LITTLE) being (1500,2000), (1500,1900), (1400,1900) and (1500,1800) without their respective measured points. Note that two of them are overlapping.
The same occurred with the \texttt{Smallpt}, \texttt{LavaMD}, \texttt{x264} and \texttt{particle Filter} applications (see Figures~\ref{fig:validation2}d,e,g,h).
\unskip

Notice that all applications except the \texttt{Particle Filter} and \texttt{kmeans} (see Figures~\ref{fig:validation2}g,f), have a point when the measured energy consumed is higher than the estimated. A vertical line represents at which time and big cluster frequency this occurs (see Figure~\ref{fig:validation2}). 
As the energy is also influenced by the performance modeling, the measured execution time in some cases takes longer when compared with the expected performance consuming more energy.

Notice the highlighted configurations (see Figure~\ref{fig:validation2}) that have similar measured energy. Those~points have the same big frequency, and, as the LITTLE frequency scales down, the execution time increases, and so does energy consumption. Further, when the big cluster frequency decreases, there is an abrupt drop in energy consumption. For instance, this can be seen in Figure~\ref{fig:validation2}b of the \texttt{{Bodytrack}} application in points above 200 seconds. This shows that the big cores play an important role in energy consumption. 

Some applications do not have a workload as well-balanced as we expected. This results in the predicted performance not being well correlated with the measured data. Even though  \texttt{Freqmine} (see~Figure~\ref{fig:validation2}c) and \texttt{kmeans} (see Figure~\ref{fig:validation2}f) have the dynamic clause included in their parallel loops, they do not present consistent performance. Additional work will be required to understand why these applications are not workload balanced. In particular, the \texttt{x264} application (see Figure~\ref{fig:validation2}e) requires more refined modelling since video processing is much more complex, and the application phases tends to vary according to the frames from the input video. Our approach shows reasonable energy~savings, even though fixing a Pareto-optimal configuration to the whole application execution. 

Table~\ref{tab:val_errors} shows the mean absolute percentage error (MAPE) between the measured and estimated performance and energy consumption of the Pareto Frontier, including the average standard deviation of the measured data. 
The highest energy and performance errors are from \texttt{Black-Scholes} (38.92\%) and \texttt{x264} (11.10\%) applications, respectively. The lowest errors are observed in \texttt{kmeans} and \texttt{LavaMD} with 10.48\% and 1.75\% of energy and performance errors, respectively. It is important to notice that the \texttt{x264} had a reasonable accuracy, considering its complexity. On average, we achieved performance and energy errors of 5.53\% and 22.30\%, respectively. 

\begin{table}[H]
    \centering
      \caption{The average standard deviation and  mean absolute percentage error of performance and energy consumption of the estimated Pareto frontier against measured data. }
    \label{tab:val_errors}  
    \scalebox{.9}[0.9]{
        \begin{tabular}{clcccccccc}
        \toprule
        \textbf{} & \textbf{} & \multicolumn{8}{c}{\textbf{Application}} \\ \cmidrule{1-10} 
         &  & \rotatebox[origin=c]{90}{\textbf{Black-scholes}} & \rotatebox[origin=c]{90}{\textbf{Bodytrack}} & \rotatebox[origin=c]{90}{\textbf{Freqmine}} & \rotatebox[origin=c]{90}{\textbf{Smallpt}} & \rotatebox[origin=c]{90}{\textbf{x264}} & \rotatebox[origin=c]{90}{\textbf{kmeans}} & \rotatebox[origin=c]{90}{\textbf{Particle Filter}} & \rotatebox[origin=c]{90}{\textbf{LavaMD}} \\ \midrule
        \multicolumn{1}{l}{\multirow{2}{*}{\textbf{Performance}}} & \textbf{MAPE} & 4.88\% & 2.55\% & 6.8\% & 2.16\% & 11.10\% & 6.96\% & 8\% & 1.75\% \\ \cmidrule{2-10} 
        \multicolumn{1}{c}{} & \textbf{Ave. Std. Deviat.} & 2.4536 & 0.8468 & 17.534 & 0.6111 & 6.8414 & 2.4148 & 0.2099 & 0.34 \\ \cmidrule{1-10}
        \multicolumn{1}{l}{\multirow{2}{*}{\textbf{Energy}}} & \textbf{MAPE} & 38.92\% & 23.97\% & 17.49\% & 27.42\% & 19.64\% & 10.48\% & 17.96\% & 22.52\% \\ \cmidrule{2-10} 
        \multicolumn{1}{c}{} & \textbf{Ave. Std. Deviat.} & 1.2244 & 1.7626 & 11.7941 & 4.9319 & 1.9157 & 7.2793 & 0.6401 & 1.1927 \\ \bottomrule
    \end{tabular}%
    }
  
\end{table}

It is important to recall that our concern is the trend of the measured Pareto frontier. Therefore, even without an exact match between measured and estimated values, the Pareto configurations are reasonable alternatives that give each application suitable trade-off choices. In the next section, we~will compare our approach against the Linux governors.

\subsection{Comparison against DVFS Governors}

All applications were executed under the \texttt{performance}, \texttt{ondemand}, and \texttt{powersave} Linux governors (see Figure~\ref{fig:validation2}).  The \texttt{performance} and \texttt{powersave} governors set the CPU statically to the highest and lowest frequency, respectively, within the borders of available minimum and maximum~frequencies. Moreover, the \texttt{ondemand} governor scales the frequency dynamically according to the current load. It boosts to the highest frequency and then likely decreases as the idle time raises. We compared our approach to these three governors as they provide a consistent variety of optimization options and are implemented on numerous mobile devices, making them competitive baselines. 

The green line in Figure~\ref{fig:validation2} shows the measured points that have higher performance and less energy consumption when compared with the \texttt{powersave} governor to all applications. Notice that there are many configurations pointed from the Pareto frontier better than the \texttt{powersave} governor. The blue and yellow lines show the measured points that have higher or similar performance when compared with the \texttt{ondemand} and \texttt{performance} governor, respectively. Notice that there are few configurations at the Pareto frontier better or similar to those governors. 

\subsubsection{Normalized Comparison}

\begin{figure}[H]
 \centering
	\includegraphics[scale=0.35]{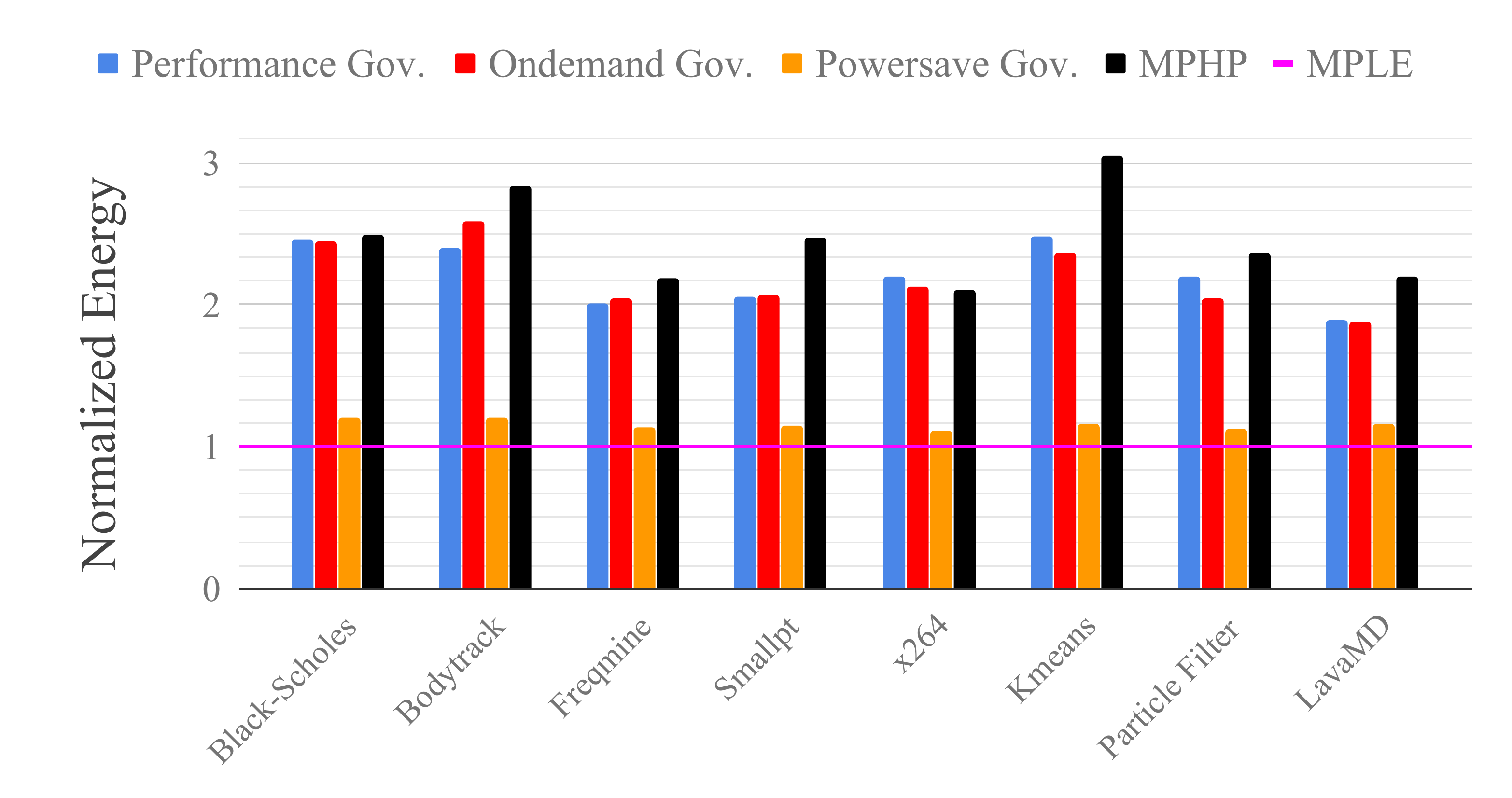}
  \caption{Normalized energy consumption by the measured Pareto configuration that provides the least energy consumption (MPLE) for each application.}	
	\label{fig:powersave}
\end{figure}

\begin{figure}[H]
 \centering
	\includegraphics[scale=0.35]{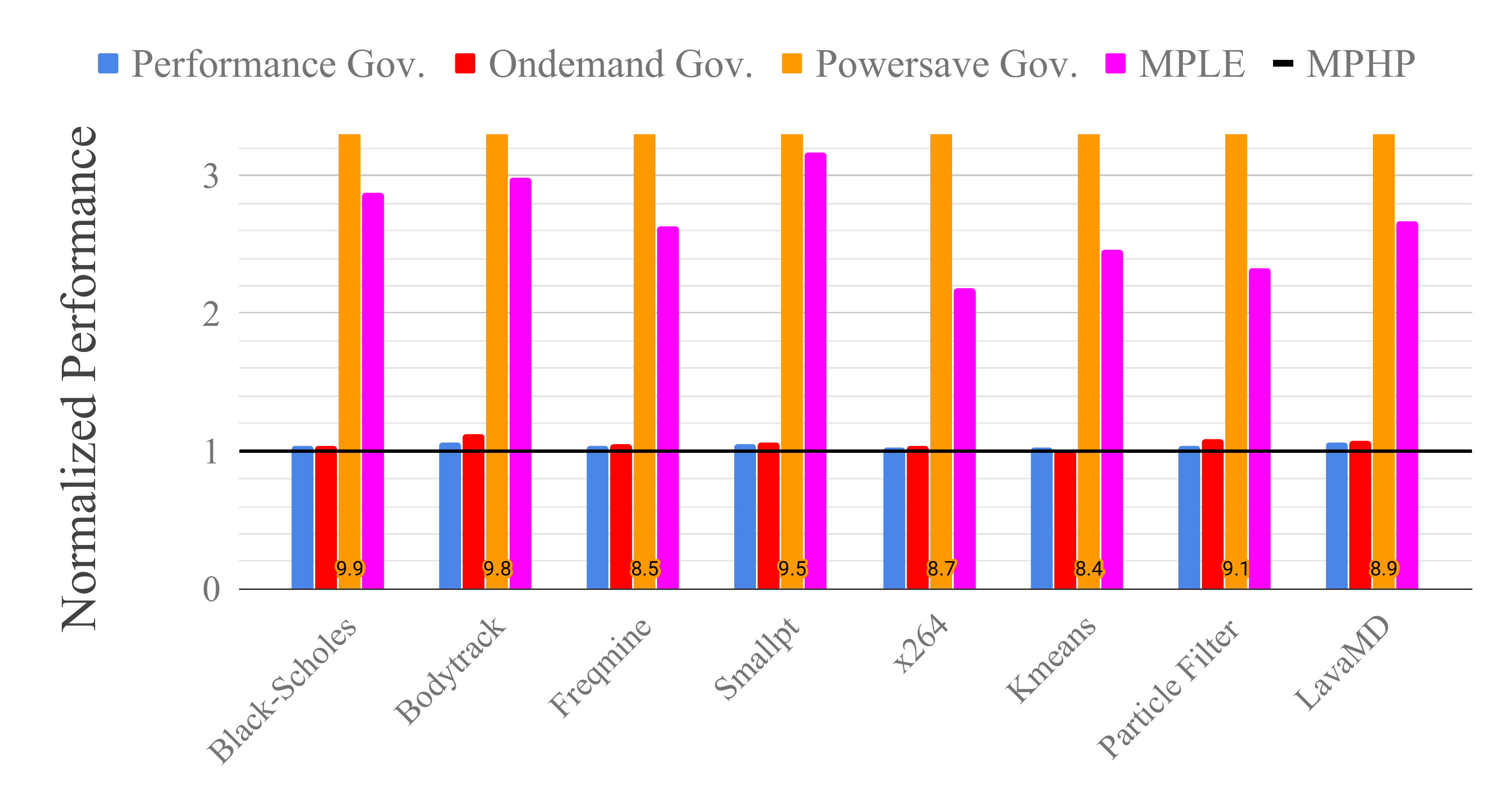}
  \caption{Normalized performance by the measured Pareto configuration that provides the  highest performance (MPHP) for each application.}	
	\label{fig:perfor}
\end{figure}

Figures~\ref{fig:powersave} and~\ref{fig:perfor} show, respectively, the normalized energy consumption and the normalized performance of all the benchmarks compared with the \texttt{performance}, \texttt{ondemand} and \texttt{powersave}~governors. We chose the measured Pareto configurations that give the least energy consumption (MPLE) and the highest performance (MPHP) to normalize the energy and the~performance, respectively, for each application.

\unskip
Figure~\ref{fig:powersave} shows that the MPLE saved more energy when compared to all governors for every~application. Also, we can observe that MPHP saved energy when compared to the \texttt{performance} and \texttt{ondemand} governors for the \texttt{x264} application. Figure~\ref{fig:perfor} shows that MPHP has a small speedup when compared to \texttt{performance} and \texttt{ondemand} Linux governors. On average, notice that MPLE has 7$\times$ higher speedup when compared with the \texttt{powersave} governor and it is only 2$\times$ slower when compared to the \texttt{performance} governor.

Table~\ref{tab:percentagegovern} shows the percentage of performance gains and energy savings when compared to the MPHP and MPLE, respectively, for all applications. Our methodology saved, on average, 54.38\%, 53.99\% and 13.67\% of energy w.r.t.\ the \texttt{performance}, \texttt{ondemand}, and \texttt{powersave} Linux governors, respectively. Also, we observed 4.23\%, 5.52\% and 89.03\%  of speedup w.r.t.\ the \texttt{performance}, \texttt{ondemand}, and \texttt{powersave} Linux governor, respectively. 

\begin{table}[H]
\centering

 \caption{Percentage of performance gains and energy savings when compared to the measured Pareto configurations that give the highest performance (MPHP) and the least energy consumption (MPLE), respectively, for all applications.}
    \label{tab:percentagegovern}
        \begin{tabular}{llccc}
            \toprule
             & \multicolumn{1}{c}{} & \multicolumn{3}{c}{\textbf{Linux Governors}} \\ \cline{1-5} 
             & \multicolumn{1}{c}{} & \textbf{Performance} & \textbf{Ondemand} & \textbf{\texttt{Powersave}} \\ \midrule
            \multicolumn{1}{l}{\multirow{2}{*}{\textbf{\texttt{Black-Scholes}}}} & {MPHP} & 3.69\% & 3.37\% & 89.96\% \\ \cline{2-5} 
            \multicolumn{1}{c}{} & {MPLE} & 59.33\% & 59.19\% & 17.70\% \\ \hline
            \multicolumn{1}{l}{\multirow{2}{*}{\textbf{\texttt{Bodytrack}}}} & {MPHP} & 6.42\% & 10.63\% & 89.94\% \\ \cline{2-5} 
            \multicolumn{1}{c}{} & {MPLE} & 58.29\% & 61.38\% & 17.15\% \\ \hline
            \multicolumn{1}{l}{\multirow{2}{*}{\textbf{\texttt{Freqmine}}}} & {MPHP} & 3.17\% & 5.24\% & 88.39\% \\ \cline{2-5} 
            \multicolumn{1}{c}{} & {MPLE} & 50.11\% & 51.15\% & 12.40\% \\ \hline
            \multicolumn{1}{l}{\multirow{2}{*}{\textbf{\texttt{Smallpt}}}} & {MPHP} & 4.53\% & 5.93\% & 89.54\% \\ \cline{2-5} 
            \multicolumn{1}{c}{} & {MPLE} & 51.30\% & 51.68\% & 12.75\% \\ \hline
            \multicolumn{1}{l}{\multirow{2}{*}{\textbf{\texttt{x264}}}} & {MPHP} & 2.67\% & 4.07\% & 88.48\% \\ \cline{2-5} 
            \multicolumn{1}{c}{} & {MPLE} & 54.62\% & 52.88\% & 10.63\% \\ \hline
            \multicolumn{1}{l}{\multirow{2}{*}{\textbf{\texttt{kmeans}}}} & {MPHP} & 2.86\% & $-$0.01\% & 88.10\% \\ \cline{2-5} 
            \multicolumn{1}{c}{} & {MPLE} & 59.77\% & 57.67\% & 14.07\% \\ \hline
            \multicolumn{1}{l}{\multirow{2}{*}{\textbf{\texttt{Particle Filter}}}} & {MPHP} & 4.18\% & 8.05\% & 89.09\% \\ \cline{2-5} 
            \multicolumn{1}{c}{} & {MPLE} & 54.54\% & 51.09\% & 10.67\% \\ \hline
            \multicolumn{1}{l}{\multirow{2}{*}{\textbf{\texttt{LavaMD}}}} & {MPHP} & 6.29\% & 6.86\% & 88.78\% \\ \cline{2-5} 
            \multicolumn{1}{c}{} & {MPLE} & 47.05\% & 46.86\% & 14.02\% \\ \hline
            \multicolumn{1}{l}{\multirow{2}{*}{\textbf{Average}}} & {MPHP} & 4.23\% & 5.52\% & 89.03\% \\ \cline{2-5} 
            \multicolumn{1}{c}{} & {MPLE} & 54.38\% & 53.99\% & 13.67\% \\ \hline
        \end{tabular}
   
\end{table}

\section{Conclusions and Future Work}
\label{sec:conclusions}
We presented a novel methodology to estimate optimal performance and energy trade-off~configurations for parallel applications on Heterogeneous Multi-Processing (HMP) systems. We devised an analytical low-overhead straightforward performance model for a given multi-thread application and an analytical application-agnostic power model for a specific two-cluster HMP system. These models, when combined, generate~the energy model which can assess all available configurations to predict an application's energy consumption. The Pareto frontier uses the provided offline performance and energy models to~select, from all available options, the optimal performance-energy trade-off.

We validated our methodology on an ODROID-XU3 board which we used to fit our models and to validate the Pareto frontier configurations. Moreover, we compared the measured Pareto frontier with the \texttt{performance}, \texttt{powersave} and \texttt{ondemand} Linux governors. Our approach achieved a reduction of the configuration search space of approximately 99\%, significantly decreasing the number of options available to identify an optimal performance-energy trade-off configuration. On average, the~performance and energy absolute percentage errors between the measured and the estimated Pareto frontier for all applications are 5.53\% and 22.30\%, respectively. The average variation of performance and energy concerning all applications are 84.25\% and 59.68\%, respectively. Also, we obtained 13.67\% of energy savings regarding the powersave governor with higher or similar performance.  These results encourage future research in performance- and energy-aware schedulers using our methodology. Furthermore, we can apply our models to predict optimal energy and performance trade-offs.

The proposed approach can be used to run parallel applications that have previously been characterized, minimizing the overhead and energy waste associated with runtime characterization. Also, offline characterization can be made as precise as necessary since resource limitation is often not an issue.
As future work, we intend to reduce the performance restrictions resulting from data-content dependent workloads, the size of the problem's input and application phase changing. Furthermore, the accuracy of the power prediction may be remodelled, considering the chip voltage as a non-linear relationship with the maximum operating frequency. 


\vspace{6pt} 



\authorcontributions{Funding acquisition, K.E. and S.X.d.S.; Project administration, K.E. and S.X.d.S.; Software, D.A.M.C.; Supervision, S.X.d.S. and K.E.; Validation, D.A.M.C.; Writing---original draft, D.A.M.C.; Writing---review and editing, D.D.S., A.F.L., K.G., J.N.Y., K.E. and S.X.d.S. All authors have read and agreed to the published version of the manuscript.
}

\funding{This work was financed in part by the Coordenação de Aperfeiçoamento de Pessoal de Nível Superior-Brasil (CAPES)-Finance Code 001, and in part by the Royal Society-Newton Advanced Fellowship award no.\ NA160108. It is also supported in part by the European Union's Horizon 2020 Research and Innovation Programme under grant agreement no.\ 779,882, TeamPlay (Time, Energy and security Analysis for Multi/Many-core heterogeneous PLAtforms).}

\acknowledgments{The authors would like to thank the anonymous reviewers.}

\conflictsofinterest{The authors declare no conflict of interest.} 



\appendixtitles{no} 



\reftitle{References}


\externalbibliography{yes}


\end{document}